\documentclass{aa}  
\usepackage{graphicx}
\usepackage{txfonts}
\usepackage{lipsum}
\usepackage[utf8]{inputenc}
\usepackage{lscape}
\usepackage{longtable}
\usepackage{booktabs}
\usepackage{placeins}
\usepackage{amsmath}
\usepackage{amssymb}
\usepackage{xcolor}
\usepackage{url}
\usepackage{multirow}
\usepackage{threeparttable}
\usepackage{soul}
\usepackage{float}
\usepackage{tablefootnote}
\usepackage{appendix}
\usepackage{orcidlink}
\usepackage{capt-of}
\usepackage{natbib}
\bibpunct{(}{)}{;}{a}{}{,} 
\usepackage{hyperref}
\hypersetup{colorlinks=true, allcolors=black, citecolor=blue}

\usepackage[
    separate-uncertainty=true,
    multi-part-units=single,
    range-phrase=--,
    range-units=single
]{siunitx}

\newcommand{\Msun}{M$_{\odot}$\xspace}

\newcommand{\Mjup}{M$_{\rm{J}}$\xspace}

\colorlet{draftcolor}{black}

\begin{document}

   \title{Population synthesis of substellar companions surviving common-envelope evolution}
   \author{Nicolás F. Miranda\inst{1}\corrauth{nmiranda@usm.cl}
        \and Alex Mauch-Soriano\inst{1,2}\email{xx}
        \and Joaquín A. Barraza-Jorquera\inst{1,2}\email{xx}
        \and Matthias R. Schreiber\inst{1}\email{xx}
        \and \\Maria Paula Ronco \inst{3}\email{xx}
        \and Felipe Lagos-Vilches\inst{1}\email{xxx}
        }

   \institute{Departamento de F\'{\i}sica, Universidad T\'ecnica Federico Santa Mar\'{\i}a, Av. Espa\~na 1680, Valpara\'{\i}so, Chile
   \and Instituto de F\'{\i}sica, Pontificia Universidad Católica de Valpara\'{\i}so, Casilla 4059, Valpara\'{\i}so, Chile
   \and Instituto de Astrof\'{\i}sica de La Plata, CCT La Plata-CONICET-UNLP, Paseo del Bosque S/N (1900), La Plata, Argentina}

   \date{Received XX / Accepted XX}

  \abstract
{Predicting the population of substellar companions around white dwarfs (WDs) requires accounting for both companions that avoid engulfment during stellar evolution and those that survive common-envelope (CE) evolution. While the former population was investigated in our previous population-synthesis study, a comprehensive population synthesis of the latter remains lacking.}
{We complement our previous study by investigating the fate of engulfed substellar companions within the same population-synthesis framework. We aim to predict the occurrence rates and properties of companions surviving CE evolution, identify the conditions permitting survival, and assess their observational detectability.}
{We extended our previous population synthesis of substellar companions around WDs, which accounts for the Galactic distributions of progenitor masses, ages, and metallicities, as well as observationally constrained initial companion distributions. Combining detailed stellar evolution with orbital evolution, we determined the stellar structure at the onset of CE evolution for $295$ engulfed companions with masses of $5$--$30$~\Mjup~drawn from a population of 20\,315 WD progenitors. We calculated CE outcomes for ejection efficiencies of $0.3$ and $1$ considering gravitational and thermal energy or the full internal energy, including recombination energy. We also modeled transit and infrared-excess detectability.}
{Survival is rare and strongly depends on the evolutionary stage at engulfment. No companions entering CE evolution during the red-giant branch or early asymptotic-giant branch survive. Including gravitational and thermal energy only, we find no survivors for a CE efficiency of $0.3$ and six for the maximum adopted efficiency of $1$. Including the full internal energy increases these numbers to three and $33$, respectively. Survivors originate exclusively from the thermally pulsing asymptotic-giant branch, when stellar envelopes can become weakly bound, allowing even relatively low-mass companions to survive. The resulting systems typically have orbital periods of a few hours. We predict occurrence rates of up to $0.16\%$ among WDs, corresponding to an upper limit of $2.53$ companions within 50\,pc. Transit detection is primarily limited by orbital geometry, whereas all predicted survivors would produce a detectable $>5\sigma$ infrared excess with JWST/MIRI at $15~\mu{\rm m}$ out to 50\,pc under our adopted observing conditions.}
{Our population synthesis predicts that CE evolution produces a small but potentially observable population of close substellar companions to WDs. Survival is governed not only by companion mass but critically by the envelope structure at engulfment, particularly during thermal pulses, and by the contribution of recombination energy to envelope ejection. Mid-infrared observations of nearby WDs offer a promising means of testing this evolutionary channel but the predicted detection rates are low.}


   \maketitle
   \nolinenumbers

\section{Introduction}

Several thousand exoplanets have been discovered around main-sequence stars, and on average each star in the Milky Way probably hosts more than one planet \citep{cassanetal12-1}. Since virtually all known planet-hosting stars will eventually evolve into white dwarfs (WDs), a substantial population of planetary companions to WDs is expected. However, the survival of these companions throughout the evolution of their host stars depends strongly on their initial orbital configuration and masses.

During the red giant branch (RGB) and asymptotic giant branch (AGB) phases, stellar mass loss causes planetary orbits to expand, while tidal interactions can lead to orbital decay and eventual engulfment \citep[e.g.,][]{villaver+livio07-1,veras16-1,NordhausSpiegel2013}. Planets initially orbiting sufficiently far from their host stars are expected to avoid engulfment and survive on wider orbits around the resulting WDs. In contrast, companions at smaller separations may enter the expanding stellar envelope. Their subsequent evolution is governed by the competition between the orbital energy released during inspiral, the binding energy of the envelope, and the processes that may ultimately destroy the companion. If sufficient energy is transferred to the envelope before the companion is tidally disrupted or evaporated, common-envelope (CE) evolution may produce a close substellar companion to the emerging WD.

The possibility that planetary companions survive CE evolution is particularly interesting in light of the discovery of substellar objects orbiting WDs at separations much smaller than the maximum radii reached by their progenitor stars. Two particularly relevant examples are the eclipsing WD--brown-dwarf binaries ZTF\,J1828+2308 and ZTF\,J1230$-$2655, which host companions of $19.5\pm0.8\,M_{\rm J}$ and $<22.1\,M_{\rm J}$ in 2.7 and 5.7 h orbits, respectively \citep{Parsons2025}. Their evolutionary histories can be reconstructed through CE evolution only if engulfment occurs after the first thermal pulse, unless very low metallicities are assumed, suggesting that the lowest-mass companions may preferentially survive CE evolution during the TP-AGB phase. 

Over the last several years, one of the most prominent examples for a potential post-CE planetary companion has been WD\,1856+534\,b, a Jupiter-like planet orbiting its host WD every 1.4 days \citep{vanderburgetal20-1}, though recent atmospheric detection has significantly disfavored the CE hypothesis \citep{macdonald2026aerosols}. Other intriguing systems are WD\,J0914+1914, where accretion of planetary material has been interpreted as evidence for a close Neptune-like companion \citep{gaensickeetal19-1}---a scenario supported by recent hydrodynamical simulations \citep{2026A&A...713A.228V}---and also 
WD\,0310-688, where an infrared excess is consistent with a Jupiter-like planet within 2~au \citep{limbachetal24-1}.  

Such systems cannot have maintained their present orbits throughout the giant phases of their host stars. Proposed explanations include inward migration following WD formation and survival of CE evolution \citep{munoz+petrovic20-1,lagosetal21-1,maldonadoetal21-1}, and second-generation formation from material expelled during post-main-sequence evolution \citep{williamsetal26}. Determining whether CE evolution can produce such systems, and how frequently this should occur, is therefore essential for understanding the origin of close substellar companions to WDs.

The survival of substellar companions during CE evolution is particularly challenging because their relatively small masses limit the orbital energy available to eject the stellar envelope. Previous investigations have explored the conditions under which planets and brown dwarfs can survive engulfment, highlighting the importance of the envelope binding energy, the efficiency of envelope ejection, and additional energy sources \citep[e.g.,][]{NordhausSpiegel2013,lagosetal21-1,Chamandy-21}. However, the outcome also depends sensitively on the evolutionary stage at which engulfment occurs. The envelope mass, its internal structure, and its binding energy change considerably during stellar evolution and depend on both the initial stellar mass and metallicity. Consequently, establishing whether a given companion can survive CE evolution does not directly determine their occurrence rate among Galactic WDs.

Population-synthesis calculations provide a means of connecting the physical conditions for CE survival with the expected occurrence rates and properties of the resulting systems. \citet{andryushin+popov21-1} investigated the survival of planets during post-main-sequence evolution using a generic initial planet population. More recently, \citet{mauchsorianoetal26-1} developed a population-synthesis model based on an initial mass function, a Galactic age--metallicity relation, and observationally constrained occurrence rates and period distribution of substellar companions around giant stars. They predicted the population of planets and brown dwarfs surviving outside the stellar envelopes and identified CE evolution as an additional potential survival channel.

The population of planets surviving CE evolution has recently been investigated by \citet{gaiboretal26-1}. They explored the survival of giant planets around $1$ and $2$\,\Msun~progenitors, considering different tidal prescriptions and CE ejection efficiencies. They found that surviving planets should generally be massive and orbit very close to their WDs, while survival becomes extremely unlikely for CE efficiencies consistent with constraints from post-CE binaries. However, their calculations were restricted to two progenitor masses at solar metallicity, and the initial planetary population was based on occurrence rates derived primarily for main-sequence stars. A prediction for the Galactic population of surviving substellar companions requires accounting for the distributions of progenitor masses, stellar ages, and metallicities, together with their effects on companion occurrence and stellar evolution.

In this paper, we extend the population-synthesis framework developed by \citet{mauchsorianoetal26-1} to investigate the population of substellar companions that survive CE evolution. The parent population spans initial companion masses from $0.8$ to $30$\,\Mjup, including both giant planets and brown dwarfs, while the CE-survival analysis presented here is restricted to the 295 engulfed companions with $M_p>5\,M_{\rm J}$, whose orbital evolution we follow around progenitor stars drawn from a representative Galactic population. By accounting for the dependence of the initial companion population on stellar mass and metallicity, and calculating the evolution of the host stars and their envelopes, we aim to predict the occurrence rates and properties of post-CE substellar companions thereby estimating if CE survival can contribute significantly to the observed population of substellar companions to WDs.

\section{Substellar engulfment analysis}

\citet{mauchsorianoetal26-1} performed population synthesis of substellar companions around WDs to investigate the population of companions that are not engulfed during the giant phases of the host star and therefore orbit the emerging WD at orbital separations $\sim5$--$20$~au. To calculate the survival, they generated a representative sample of $20\,315$ WD progenitors in the mass range $1$--$4$~\Msun. Of these progenitors $1000$ hosted a substellar companion. Assuming relatively weak tides \citep{Villaver_2009,Mustill_2012}, they found that $262$ of these companions survived, which translates to an occurrence rate of just $1.3\%$\ of all simulated WD progenitors. Uncertainties related to the assumed age--metallicity relation, the strengths of stellar tides, and the initial companion distributions allow this fraction to increase to a strict upper limit of $\sim8\%$. Here we investigate the fate of the remaining $738$ companions that are engulfed during the giant-branch evolution. For the CE-survival calculations, we restricted the analysis to the $295$ engulfed companions with $M_{\rm p}>5\, M_{\rm J}$.

\subsection{Determination of the common-envelope onset}

To determine the onset of the CE phase, we combined the stellar evolutionary tracks obtained with the Modules for Experiments in Stellar Astrophysics (MESA) code (version r23.05.1;\,\citealt{jermyn2023}) with the orbital tracks computed with the algorithm called Final exoplAnet orbiTal dEstination around evolved Stars (FATES) developed by \citet{ronco2020}. These orbital tracks describe the evolution of the orbit of the substellar companions under prescriptions of stellar mass loss and tidal interactions, which can induce orbital and eccentricity changes. For each system, MESA provides the stellar mass, ${M_{\star}(t)}$, and stellar radius, $R_{\star}(t)$, whereas FATES provides the evolution of the semimajor axis, $a(t)$, and eccentricity, $e(t)$, of the substellar companion. The companion mass, $M_p$, was fixed for each system.

We determined the onset of engulfment by comparing the Roche volume radius with the radius of the host star. For eccentric systems, we assumed that Roche lobe filling occurs at periastron, where the orbital separation reaches its minimum value. This represents a reasonable assumption as the orbital periods are generally much shorter than the evolutionary timescale of giant stars. The periastron separation is given by $D_{p}(t)=a(t)[1-e(t)]$. The Roche-lobe radius of the giant was then calculated using the Eggleton approximation \citep{eggleton1983} evaluated at the periastron separation.

To determine when the star fills its Roche lobe, we combined the original stellar and orbital time samples over their common time interval. We identified the first pair of consecutive samples for which $R_\star-R_{\rm L}$ changes from negative to zero or positive. Within this interval, we numerically solved $R_\star(t)-R_{\rm L}(t)=0$, evaluating the Roche-lobe radius from the linearly interpolated quantities $a(t)$, $e(t)$, $M_\star(t)$, and $R_\star(t)$. The resulting root defined the CE-onset time, $t_{\rm CE}$, at which we evaluated $a_i=a(t_{\rm CE})$, $M_{\star}=M_{\star}(t_{\rm CE})$, and $R_{\star}=R_{\star}(t_{\rm CE})$. 

\subsection{Stellar evolution and common-envelope energy formalism}
\label{sect:ENERGY_FORMALISM}

\begin{figure*}
    \centering
    \includegraphics[width=\textwidth]{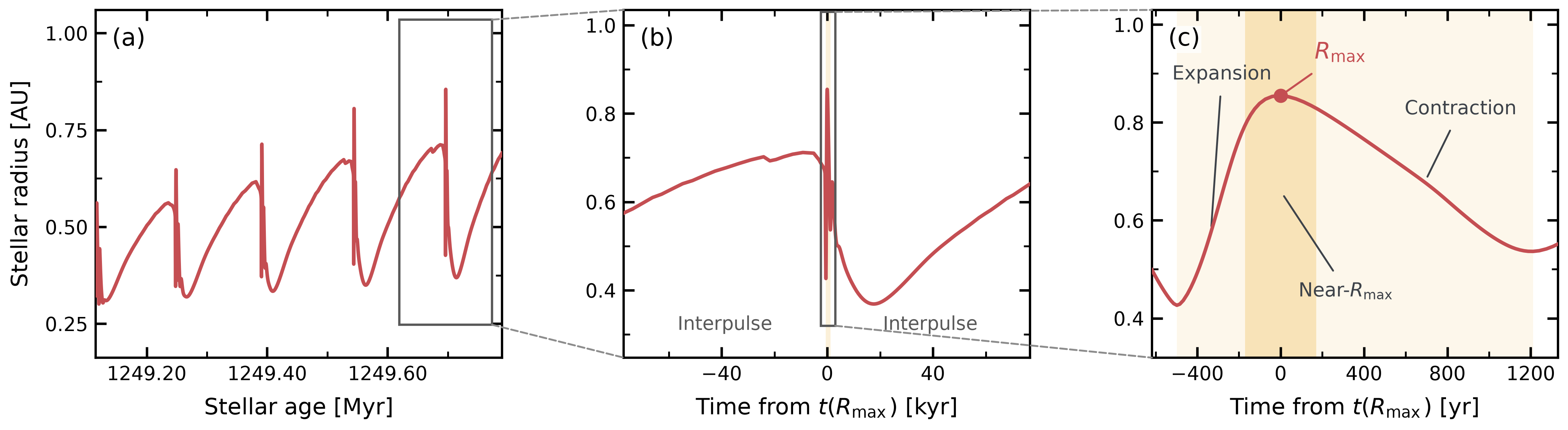}
    \caption{Stellar-radius evolution showing the thermal-pulse classification for a progenitor with initial mass $2.041\,M_\odot$ and metallicity $Z=0.0112$. Radial responses to thermal pulses during the TP-AGB (a). A zoom-in of a radial response showing the thermal cycle associated with one pulse, including the surrounding interpulse evolution (b). The expansion toward the maximum stellar radius and the later contraction (c). The light-orange region marks the radial-pulse window. The dark-orange region marks the Near-$R_{\rm max}$ interval, defined as $\pm10\%$ of the radial-window duration around $t(R_{\rm max})$.}
    \label{fig:PulsesMethodology}
\end{figure*}

We used the MESA code to simulate stellar evolution, following the physical prescriptions described by \citet{mauchsorianoetal26-1}. We adopted exponential diffusive overshooting at the outer zone of the convective core with $f_{\rm ov} = 0.016$ and $f_0 = 0.008$. During core H-burning we set the overshooting efficiency to zero for $M \leq 1.2$ \Msun, smoothly increasing with a mass-dependent scaling between $1.2 \text{ and }2.0$ \Msun and full efficiency for $M \geq 2.0$ \Msun~ as suggested by \citet{claret&torres2017}. During He-burning, we applied full overshooting for all stellar masses \citep{spruit2015,constantinoetal2015}.
At $t_{\rm{CE}}$, we recovered the stellar structure and properties from the corresponding MESA evolutionary sequence, including the H-free core mass $M_c=M_c(t_{\rm{CE}})$, and the envelope mass, $M_{\rm{env}}=M_{\star}(t_{\rm{CE}})-M_c(t_{\rm{CE}})$.
This approach assumes that the dynamical inspiral timescale is short compared with the timescale over which the stellar structure changes, even during thermal pulses, allowing us to use the stellar structure at CE onset as a fixed snapshot for the energy calculation. This approximation is motivated by the short inspiral timescales found in the literature \citep[e.g.,][]{Clayton_2017,OConnor2023}, and its application during thermal pulses is discussed in Sect.~\ref{sect:DependenceEnvelopeStructure} and shown in Fig.~\ref{fig:Stellar&OrbitalEvolution}.

For CE evolution, we adopted the energy formalism by \citet{IbenLivio1993}, as commonly implemented in population synthesis studies \citep[e.g.,][]{Zorotovic2010, Toonen2012, lagosetal21-1, Belloni2024}, where the outcome of CE evolution is approximated by the balance between the change in the orbital energy and the envelope binding energy, defined as
\begin{equation}
    -E_{\rm{bind}}=\alpha_{\rm{CE}}\Delta E_{\rm{orb}} = \alpha_{\rm{CE}}\left(\frac{G M_{c}M_{p}}{2 a_{f}}-\frac{G M_{\star} M_{p} }{2 a_i}\right),
\label{eq:EnergyEQ}
\end{equation}
where $|E_{\rm{bind}}|$ is the magnitude of the envelope binding energy. Throughout this work $E_{\rm{bind}}<0$ denotes a bound stellar envelope. $\Delta E_{\rm{orb}}$ is the orbital energy released during inspiral, $G$ is the gravitational constant, $a_{f}$ is the semimajor axis after CE ejection, and $\alpha_{\rm{CE}}$ is a CE efficiency parameter that represents the fraction of orbital energy released during orbit decay that is successfully transferred into the stellar envelope for ejection, which is estimated to be $\sim$0.24--0.41 for post-CE WD--brown-dwarf binaries \citep{Zorotovic2022}. For that reason, in this paper we adopted two values for $\alpha_{\rm{CE}}$, which are $\alpha_{\rm{CE}}=0.3$ as our fiducial efficiency and $\alpha_{\rm{CE}}=1$ as the maximum-efficiency case. If the envelope is successfully ejected, then the orbital semimajor axis of the surviving substellar companion is given by
\begin{equation}
    a_{f}=\frac{G\,M_{c}\,M_{p}}{ 2(-E_{\rm{bind}}/\alpha_{\rm{CE}}\ + GM_{\star}M_{p}/2a_i)}.
\label{eq:af}
\end{equation}

We calculated the binding energy with the prescription used by \citet{Belloni2024}, integrating over the stellar envelope from the H-free core boundary to the surface of the star. We defined the core-envelope boundary using an H mass-fraction threshold of $X_{\rm H}=0.01$ \citep{scherbak+Fuller2023}. In our model we took into account three different prescriptions. First, the gravitational binding energy of the envelope is given by
\begin{equation}
    E_{\rm{pot}}=-\int_{M_c}^{M_{\star}} \frac{Gm}{r}\, dm ,
\label{eq:Epot}
\end{equation}
where $m$ is the mass coordinate, and $r(m)$ is the radius at a given mass coordinate. This term represents the gravitational contribution to the envelope binding energy.
We also calculated the envelope binding energy accounting for the contribution of thermal energy, which reduces the net energy required to unbind the envelope, as
\begin{equation}
    E_{\rm{pot+therm}}=-\int_{M_c}^{M_{\star}} \frac{Gm}{r}\, dm + \int_{M_c}^{M_{\star}} \frac{\frac{3}{2}P_{\rm{gas}}+3P_{\rm{rad}}}{\rho}\,dm,
\label{eq:Epot+Eterm}
\end{equation}
where $P_{\rm gas}, P_{\rm rad}$, and $\rho$ are the gas pressure, radiation pressure and density, respectively. The first term on the right-hand side is the gravitational energy, while the second term on the right-hand side corresponds to the thermal energy of an ideal monatomic gas and the radiation-energy density, which reduce the required energy for ejection.

In addition, we also calculated the envelope binding energy accounting for the total specific internal energy provided by the MESA equation of state,
\begin{equation}
    E_{\rm{pot+int}}=-\int_{M_c}^{M_\star} \frac{Gm}{r}\, dm + \int_{M_c}^{M_\star} u(m)\,dm,
\label{eq:Epot+Eint}
\end{equation}
where $u(m)$ is the specific internal energy of each mass shell calculated with MESA. The second term on the right-hand side includes the full specific internal energy supplied by MESA, including ionization/recombination contributions, which reduce the net envelope binding energy if available for envelope ejection. 

In summary, in this work we calculated three definitions of the envelope binding energy, but the purely gravitational prescription was only used to characterize the envelope structure. The CE survival calculations correspond to the combination of the two adopted $\alpha_{\rm CE}$ values and the two binding-energy prescriptions given in Eqs.\,\eqref{eq:Epot+Eterm} and \eqref{eq:Epot+Eint}, being $E_{\rm pot}+E_{\rm therm}$ with $\alpha_{\rm CE}=0.3$ our fiducial model, while E$_{\rm pot}+E_{\rm int}$ with $\alpha_{\rm CE}=1$ is our most optimistic limiting case. 

To characterize the structure and binding of the stellar envelope, we calculated an effective envelope-structure parameter $\lambda$ for all energy prescriptions as
\begin{equation}
    \lambda=\frac{GM_{\star}M_{\rm{env}}}{R_{\star}|E_{\rm{bind}}|}
    \label{eq:lambda}
\end{equation}
where $E_{\rm{bind}}$ refers to gravitational, gravitational-thermal, and gravitational-internal depending on the energy prescription. In contrast to the usual $\alpha\lambda$ implementation of the CE energy formalism \citep[e.g.,][]{Webbink1984,Zorotovic2022}, we did not use $\lambda$ to calculate the final orbital separation because the energies were obtained directly from MESA and $a_f$ was calculated from $E_{\rm{bind}}$ with Eq.\,\eqref{eq:af}. The $\lambda$ values obtained in this paper were used only as structural diagnostics, providing a dimensionless measure of how strongly the envelope is bound, allowing us to compare different thermal-pulse phases.

\subsection{Evolutionary stage of the giant at CE onset}
\label{sect:evolutionarystage}

The survival of an engulfed substellar companion depends crucially on the evolutionary stage of its host star at the onset of CE evolution \citep{Casewell2024, Parsons2025}. This is particularly important during the thermally pulsing asymptotic giant branch (TP-AGB), where the stellar radius and envelope structure vary substantially over a thermal-pulse cycle \citep{Karakas2014,GonzalezBolivar2022}. As the envelope expands during a pulse, it can become less tightly bound, and especially when assuming that recombination energy contributes to envelope ejection, the conditions for survival can change significantly over relatively short evolutionary timescales \citep{Reichardt2020,GonzalezBolivar2022}. It is therefore necessary to identify the thermal-pulse phase at which CE evolution begins. The procedure used to characterize the TP-AGB evolution is illustrated in Fig.\,\ref{fig:PulsesMethodology}.

We first separated systems entering CE before the TP-AGB from those reaching contact after thermal pulses have begun. Systems entering CE during the red giant branch were classified as RGB. After core He burning had ended, systems reaching CE before the onset of the first confirmed He-shell flash were classified as early AGB (E-AGB). For systems that reached the TP-AGB, we further determined their position within the thermal-pulse cycle.

Thermal pulses were identified from the evolution of the He-burning and H-burning luminosities in the MESA tracks. We restricted the pulse search to models after core He exhaustion. Candidate He-shell flashes were first identified as peaks in $\log L_{\rm He}$ with a minimum prominence of 0.5 dex. To avoid numerical fluctuations, a candidate was retained only if He burning dominates over H burning, $\log L_{\rm He}>\log L_{\rm H}$ and the H-burning luminosity decreases by at least 0.5 dex.

Panel (a) of Fig.\,\ref{fig:PulsesMethodology} shows the repeated radial response associated with successive thermal pulses during the TP-AGB. For each confirmed pulse, we defined an associated thermal cycle. Consecutive thermal cycles were separated at the temporal midpoint between shell-flash peaks. The first cycle began at the onset of the first pulse, while the last extended to the end of the available TP-AGB track. Panel (b) illustrates one such cycle, showing that most of the cycle corresponds to interpulse evolution, while the radial response associated with the pulse occupies a much shorter time interval.

Within each thermal-pulse cycle, we defined the radial-pulse window as the interval between the radius minima surrounding $R_{\rm max}$ (panel (c) of Fig.~\ref{fig:PulsesMethodology}). Systems entering CE outside these windows were classified as Interpulse, while those entering within 10\% of the window duration around $t(R_{\rm max})$ were classified as Near-$R_{\rm max}$. The remaining systems within the window were classified as Expansion or Contraction, depending on whether the stellar radius is increasing or decreasing. 

\begin{table*}
\centering
\caption{Main results of the CE calculations for the 295 analyzed engulfed systems with $M_p>5\,M_{\rm J}$. The table lists the adopted energy prescriptions and efficiency parameters, survival fractions, final orbital properties, and transit-detectability metrics.}
\label{tab:main_results}
\small
\begin{tabular}{lccccccccc}
\hline
\noalign{\vskip 0.5mm}

\multicolumn{1}{c}{Energy Formalism} &
$\alpha_{\rm CE}$ &
$f_{\rm survival}$ &
$f_{\rm CE}$ &
\(a_f/a_{\rm RLOF}\) &
$\tilde{P}_f$&
$\tilde{M}_{p}$ &
$\langle p_{\rm transit}\rangle$ &
$\langle p_{\rm detection}^{120\,{\rm s}}\rangle$\\
&
&
&
&
&
(h) &
($M_{\rm J}$) &
(\%) &
(\%) \\
\hline
\noalign{\vskip 0.5mm}
$E_{\rm pot}+E_{\rm therm}$
& 0.3
& $0/295$ (0.0\%)
& $0/20315$ (0)
& \ldots
& \ldots
& \ldots
& \ldots
& \ldots \\

$E_{\rm pot}+E_{\rm therm}$
& 1.0
& $6/295$ (2.0\%)
& $6/20315$ \((2.95\times10^{-4})\)
& 1.04–1.70
& 2.69
& 27.09
& 13.44
& 13.03 \\

$E_{\rm pot}+E_{\rm int}$
& 0.3
& $3/295$ (1.0\%)
& $3/20315$ \((1.48\times10^{-4})\)
& 1.05–2.69
& 7.60
& 25.66
& 9.95
& 9.66 \\

$E_{\rm pot}+E_{\rm int}$
& 1.0
& $33/295$ (11.2\%)
& $33/20315$ \((1.62\times10^{-3})\)
& 1.06–8.79
& 4.50
& 22.94
& 10.04
& 9.76 \\
\hline
\end{tabular}
\end{table*}

\subsection{Substellar companion survival criterion}
\label{sect:planetarysurvival}

Several criteria for the survival of planets or substellar companions have been discussed in the literature. The thermal disruption criterion proposed by \citet{LivioSoker1984} has been frequently used to evaluate the survival of substellar companions  \citep[e.g.,][]{Nelemans-98,SiessLivio1999b,lagosetal21-1}. This criterion is based on the assumption that an engulfed substellar body is thermally disrupted when the local sound speed exceeds the body's surface escape velocity. \citet{OConnor2023} revisited this criterion and found that the heating time of the substellar companion is significantly longer than the predicted inspiral timescale and thermal disruption according to \citet{LivioSoker1984} is therefore unlikely to fully destroy engulfed companions.   
We therefore did not consider thermal disruption as a survival criterion. \citet{OConnor2023} argued that ram-pressure disruption and ablation are negligible for giant planets because their densities substantially exceed those of the surrounding stellar envelope. However, hydrodynamical studies suggest that this conclusion does not necessarily hold under all conditions. \citet{JiaSpruit2018} showed that sufficiently strong ram pressure can deform and ultimately disrupt an engulfed planet, while \citet{Yarza2023} found that ram-pressure disruption can dominate for low-mass substellar companions engulfed during the early RGB. More recently, \citet{Lau2025} performed hydrodynamical simulations resolving the gaseous structure of a Jupiter-mass planet during engulfment and found substantial, continuous mass loss through ablation prior to its destruction. Extending this work, \citet{Lau2026} identified Kelvin--Helmholtz instabilities at the planetary surface as the primary mechanism driving this gradual mass loss.

These recent hydrodynamical simulations, however, are focused on Jupiter-mass planets, whereas the companions surviving CE evolution in our models are generally more massive. Their deeper gravitational potential wells may make them less susceptible to ablation, particularly within extended giant envelopes where gravitational drag can become increasingly important \citep{Lau2026}. Nevertheless, dedicated hydrodynamical simulations covering the brown-dwarf mass regime are required to quantify the importance of ablation for these objects. We therefore neglected gradual ablation as an additional destruction mechanism in our population synthesis, while acknowledging that this assumption may lead to an overestimate of the predicted survival fraction.

Finally, we calculated the critical semimajor axis at which the substellar companion fills its Roche lobe, initiating Roche-lobe overflow (RLOF), $a_{\rm RLOF}$, using the \citet{eggleton1983} approximation assuming a circular post-CE orbit, with mass ratio $q=M_p/M_c$ and imposing $R_{\rm L,p}=R_p$, for a fixed $R_p=1\,R_{\rm J}$. This reference radius is motivated by evolutionary models predicting radii close to Jupiter's radius for mature giant planets and brown dwarfs in our adopted mass range \citep{Baraffe2003}. The sensitivity of the survival criterion to this assumption is summarized in Appendix~\ref{app:substellar_radius}. If the stellar envelope remains present at this point, continued orbital angular-momentum loss through interaction with the envelope may render the mass transfer unstable and prevent the companion from surviving. We did not model this subsequent evolution explicitly and therefore adopted companion Roche-lobe overflow as a terminal condition.

\section{Predicted companion distributions and occurrence rates}

The survival of substellar companions during CE evolution depends on the orbital energy available to eject the stellar envelope, which at fixed stellar and orbital properties increases with companion mass. As introduced in Sect.~2, we analyze 295 engulfed companions with $M_{\rm p}>5\,M_{\rm J}$, selected from the parent sample of 738 engulfed systems. We defined the number of surviving systems in each CE model as $N_{\rm surv}$, the fraction of survivors as $f_{\rm survival}=N_{\rm surv}/295$ and $f_{\rm CE}=N_{\rm surv}/20315$ as the fraction of WDs predicted to host a substellar companion that survived CE evolution. We investigate four combinations, that is $\alpha_{\rm CE}=0.3$ and 1, both with the $E_{\rm pot}+E_{\rm therm}$ and $E_{\rm pot}+E_{\rm int}$ binding-energy prescriptions.

Table~\ref{tab:main_results} summarizes the global CE outcomes and properties of the surviving systems. In the fiducial $E_{\rm pot}+E_{\rm therm}$ with $\alpha_{\rm CE}=0.3$ model, none of the analyzed companions survive, while increasing the efficiency to $\alpha_{\rm CE}=1$ produces six survivors. Thus, even at the maximum adopted efficiency, CE survival remains very rare under the $E_{\rm pot}+E_{\rm therm}$ prescription.

Adopting the full internal-energy prescription, $E_{\rm pot}+E_{\rm int}$, which includes recombination energy, substantially changes the outcome. For $\alpha_{\rm CE}=0.3$, three companions survive, demonstrating that envelope ejection can become energetically possible even for relatively inefficient CE evolution. For $\alpha_{\rm CE}=1$, the number of survivors increases to 33. In particular, the increase from six to 33 survivors at $\alpha_{\rm CE}=1$ illustrates the importance of the additional energy available when the full internal energy, including recombination energy, is adopted. In what follows we describe how survival depends on the envelope structure at the onset of CE evolution and the initial parameters before we estimate the fraction of WDs with detectable substellar companions that survived CE evolution.

\subsection{Dependence on envelope structure}
\label{sect:DependenceEnvelopeStructure}

\begin{figure*}
    \centering
    \includegraphics[width=0.9\textwidth]{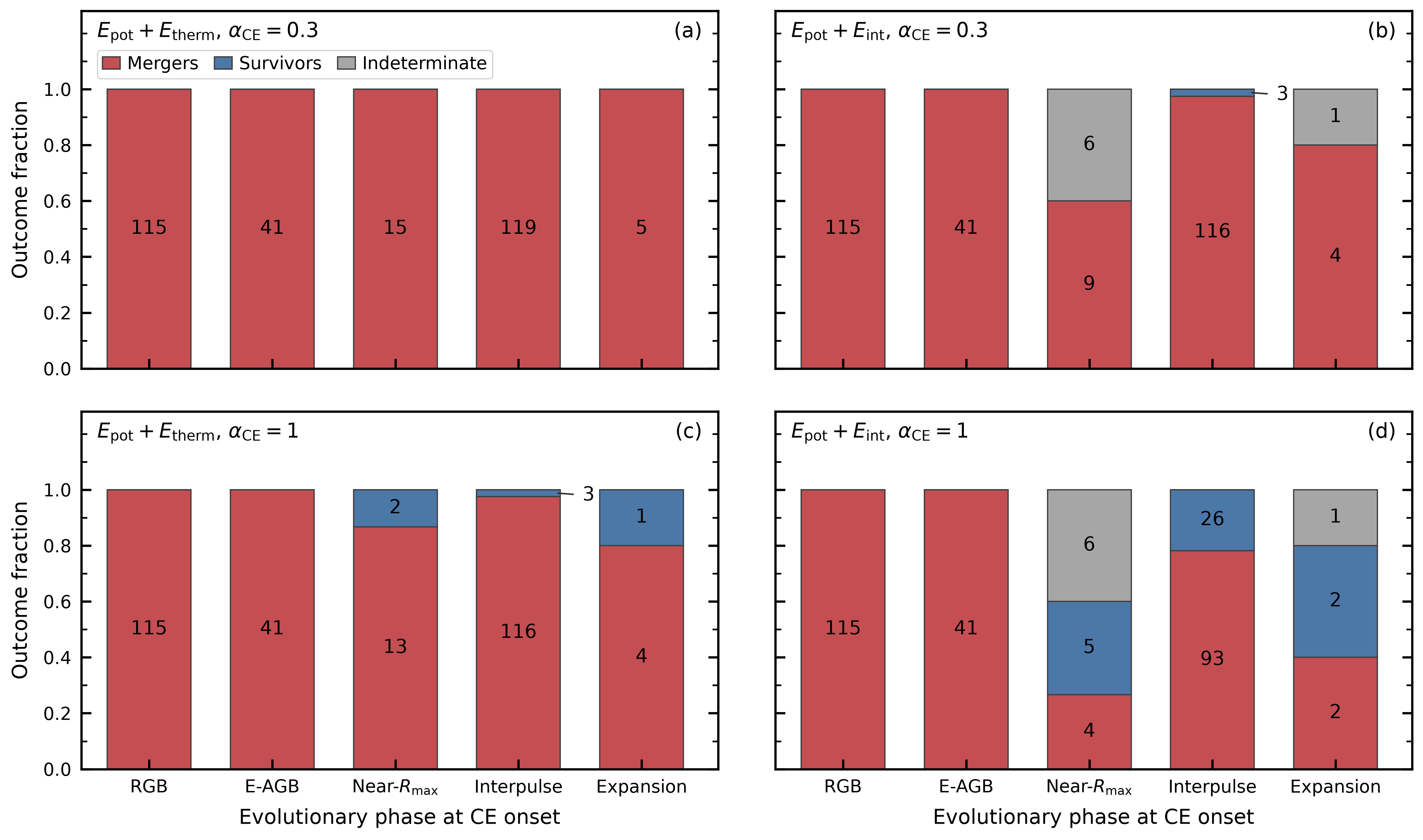}
    \caption{CE outcomes as a function of evolutionary stage at CE onset for the 295 analyzed systems. Panels (a) and (b) show the $E_{\rm pot}+E_{\rm therm}$ and $E_{\rm pot}+E_{\rm int}$ prescriptions, respectively, for $\alpha_{\rm CE}=0.3$. Panels (c) and (d) show the corresponding prescriptions for $\alpha_{\rm CE}=1$. Red, blue, and gray regions represent mergers, survivors, and indeterminate systems with $E_{\rm bind}\geq0$, respectively. While more systems survive when recombination energy is included, even without this additional energy, some planets can survive if CE evolution starts with the host star on the late AGB.}
    \label{fig:Pulses_Histogram}
\end{figure*}

\begin{figure*}
    \centering
    \includegraphics[width=0.9\textwidth]{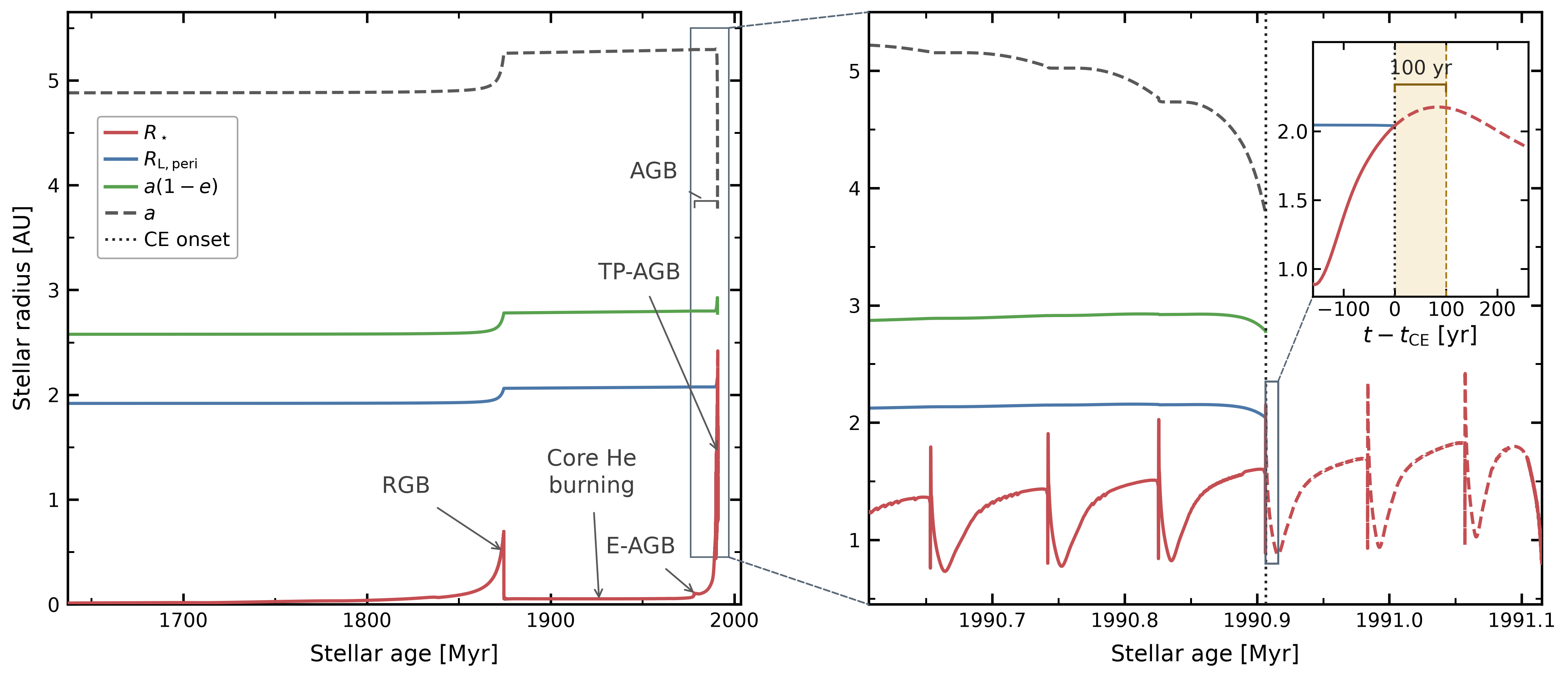}
    \caption{Orbital and stellar evolution of a 9.39\,\Mjup survivor that enters CE evolution near the maximum stellar radius under the $E_{\rm pot}+E_{\rm therm}$ prescription with $\alpha_{\rm CE}=1$. The left panel shows the host-star evolution from the RGB through the AGB, while the right panel shows a zoom-in of the TP-AGB pulses around CE onset, with the contact region enlarged in the inset. Red, blue, and green curves show the stellar radius, the periastron stellar Roche-lobe radius, and the orbital periastron distance, respectively. Orbital tracks terminate at contact, therefore the dashed red curve is the continuation of the undisturbed stellar track. The shaded orange 100 yr interval shows the fixed-structure approximation adopted in Sect.~\ref{sect:ENERGY_FORMALISM}, rather than a CE duration calculated for this system. The survivor shows the importance of the evolutionary stage at CE onset, in particular during a thermal pulse where even a low-mass companion can survive without recombination energy.}
    \label{fig:Stellar&OrbitalEvolution}
\end{figure*}

We found that the evolutionary state of the giant star has a strong impact on survival as it determines the envelope binding energy and therefore how much orbital energy is required to expel the envelope. In particular, we discovered peculiar cases that indicate a contradiction between the CE energy formalism and MESA. When the full internal-energy prescription is adopted, including recombination energy, we found cases with $E_{ \rm{bind}}\ge0$. The occurrence of such non-negative envelope binding energies highlights an inconsistency between the condition for terminating AGB evolution in MESA, which is given by a lower limit on the envelope mass, and the energy formalism used for CE evolution.

A positive total envelope energy may indicate that the envelope is approaching a state in which additional processes, including mass loss or dynamical expansion, become important. Whether the envelope would actually be ejected before or during the interaction with the companion cannot be established from its integrated energy alone. It might also well be that the appearance of such cases simply illustrates that including large fractions of recombination energy is not realistic. We call cases with a positive binding energy at the onset of CE evolution indeterminate as the CE energy formalism does not allow us to determine the final orbital separation. Therefore, as the basic idea of energy conservation breaks down in these cases, we did not count these indeterminate systems as survivors of CE evolution, but they remain in the total sample of 295 modeled systems, recognizing that their subsequent evolution requires a more detailed treatment of the envelope dynamics including possible energy losses through efficient radiative cooling \citep{OConnor2023}.

Figure~\ref{fig:Pulses_Histogram} shows the outcome of evolving all engulfed companions with masses exceeding $5$\,\Mjup~as a function of the evolutionary stage of the host star at engulfment. Of the 295 systems, 115 enter CE evolution during the RGB and another 41 during the E-AGB. None of these systems survive under any adopted prescriptions. At these evolutionary stages, the stellar envelopes remain sufficiently tightly bound that the orbital energy released by the engulfed companions is insufficient to eject them before the companions fill their Roche lobes. No system enters CE during the contraction phase, so this category is omitted from Fig.~\ref{fig:Pulses_Histogram}.

For $E_{\rm pot}+E_{\rm therm}$ with $\alpha_{\rm CE}=1$, two companions survive near the maximum stellar radius, three during the Interpulse phase, and one during the Expansion phase. Using the $E_{\rm pot}+E_{\rm int}$ prescription substantially increases the number of survivors, particularly during Interpulse evolution, while also increasing the number of indeterminate systems with $E_{\rm bind}\ge0$ from zero to seven. Although most survivors occur during Interpulse evolution in absolute numbers, for $E_{\rm pot}+E_{\rm int}$ with $\alpha_{\rm CE}=1$ the survival fractions in each phase are 5/15 ($33.3^{+25.0}_{-18.2}\%$) for Near-$R_{\max}$, 26/119 ($21.8^{+8.2}_{-6.5}$\%) for Interpulse, and 2/5 ($40.0^{+36.9}_{-28.2}$\%) for Expansion, where the uncertainties denote 95\% Wilson confidence intervals \citep{Wilson1927}. These results demonstrate that the evolutionary stage at which CE evolution begins is crucial for determining whether a substellar companion survives. In particular, the reduced envelope binding energies reached during the TP-AGB, combined with the contribution of recombination energy, can allow envelope ejection under conditions for which survival would otherwise be impossible.

The importance of stellar structure at CE onset is particularly evident for the few relatively low-mass companions that survive under the $E_{\rm pot}+E_{\rm therm}$ prescription with $\alpha_{\rm CE}=1$. These systems enter CE evolution close to the maximum stellar radius reached during a thermal pulse, when their envelopes are weakly bound, with $\lambda_{\rm therm}\simeq0.73$--$0.94$. Under these conditions, even a companion with a relatively small orbital-energy reservoir can eject the envelope before filling its Roche lobe. This demonstrates that the timing of engulfment relative to the thermal-pulse cycle can be crucial for survival, allowing some lower-mass companions to survive under otherwise unfavorable conditions.

An example of such an evolution is given in Fig.~\ref{fig:Stellar&OrbitalEvolution}. The companion avoids engulfment during earlier evolutionary phases and enters CE evolution during the expansion associated with a thermal pulse, when the stellar envelope is weakly bound. Importantly, the radial evolution during thermal pulses occurs on timescales of hundreds to thousands of years, considerably longer than the short inspiral timescales expected for CE evolution \citep[e.g.,][]{OConnor2023}. We therefore approximate the stellar structure during the inspiral by the MESA model at $t_{\rm CE}$, as described in Sect.~\ref{sect:ENERGY_FORMALISM}.

The $9.39\,M_{\rm J}$ survivor shown in Fig.~\ref{fig:Stellar&OrbitalEvolution} has a $\lambda_{\rm therm}\simeq0.94$ at CE onset, whereas another survivor of $29.16\,M_{\rm J}$ under the same energy prescription and efficiency has $\lambda_{\rm therm}\simeq0.25$. These examples show that survival depends on both the envelope structure and the available orbital energy. The total stellar mass, envelope mass, and radius at engulfment set the binding-energy scale, as shown in Eq.~\eqref{eq:lambda}, while the orbital-energy change depends on the stellar and core masses, the companion mass, and the change in orbital separation from $a_i$ to $a_f$, as shown in Eq.~\eqref{eq:EnergyEQ}. In the system shown in Fig.~\ref{fig:Stellar&OrbitalEvolution}, the large stellar radius at engulfment ($R_\star=2.04\, {\rm au}$) reduces the binding-energy scale, compensating for the smaller companion mass. Thus, survival depends on the combination of envelope structure and available orbital energy, rather than on a universal value of $\lambda$.

\subsection{Dependence on initial parameters}
\label{sect:DependenceOnInitialParameters}

\begin{figure*}
    \centering
    \includegraphics[width=0.9\linewidth]{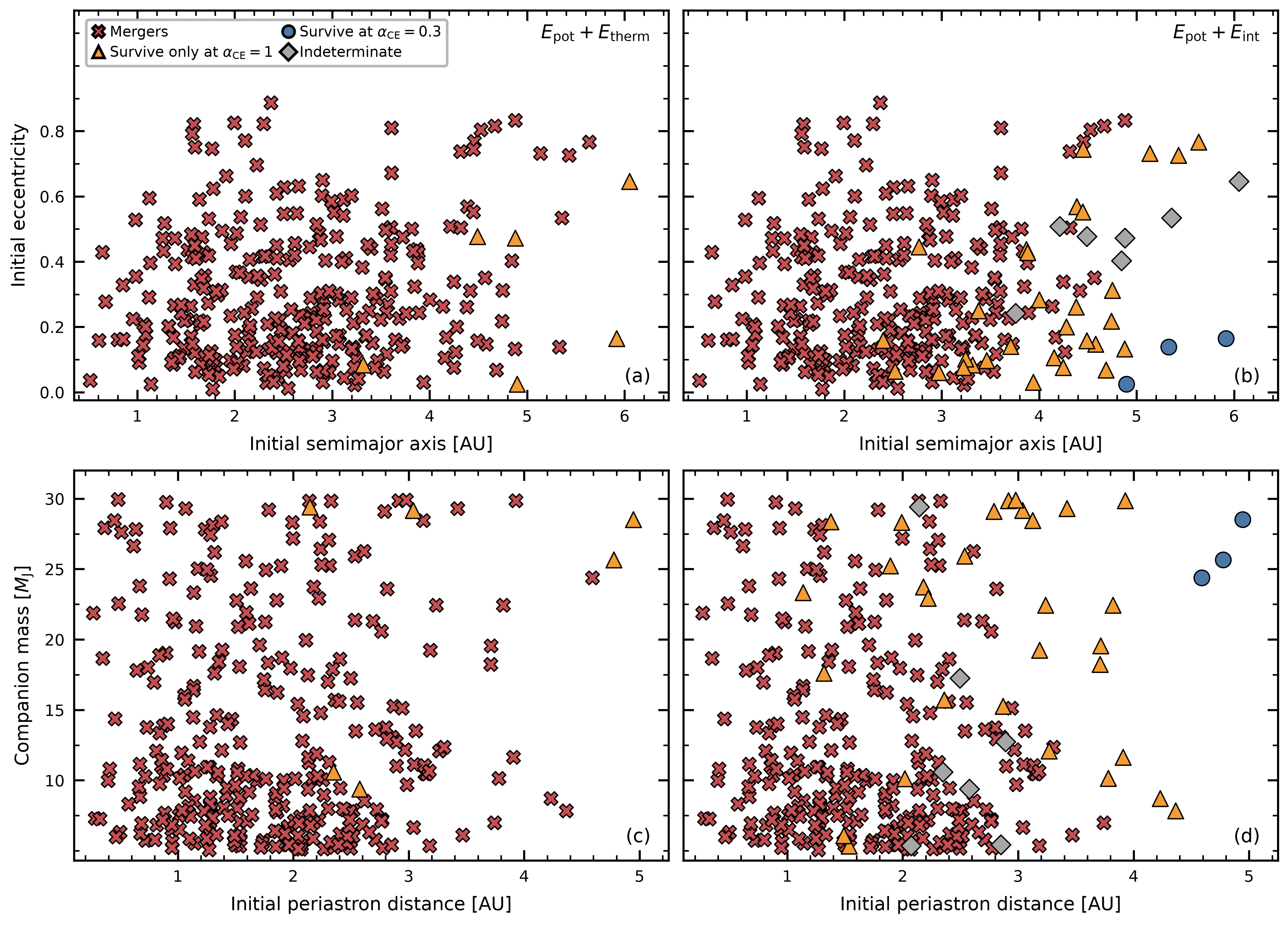}
    \caption{CE outcomes as a function of initial conditions of the 295 systems. Panels (a) and (b) show the initial eccentricity versus the initial orbital separation in au, for the $E_{\rm pot}+E_{\rm therm}$ and $E_{\rm pot}+E_{\rm int}$ prescriptions, respectively. Panels (c) and (d) show companion mass versus the initial periastron distance for the same two prescriptions. Red crosses indicate mergers at both adopted CE efficiencies. Orange triangles represent survival only for $\alpha_{\rm CE}=1$. Blue circles indicate survival at $\alpha_{\rm CE}=0.3$. Gray diamonds represent indeterminate systems with $E_{\rm bind}\geq0$. No single property determines survival alone. While wider initial orbits and higher companion masses favor survival, even without including recombination energy, planets with masses as low as $\sim9.39$~\Mjup can survive if CE evolution starts when the host star is close to the maximum radius during a thermal pulse.}
    \label{fig:InitialConditionsOutcomes}
\end{figure*}

\begin{figure*}
    \centering
    \includegraphics[width=0.9\textwidth]{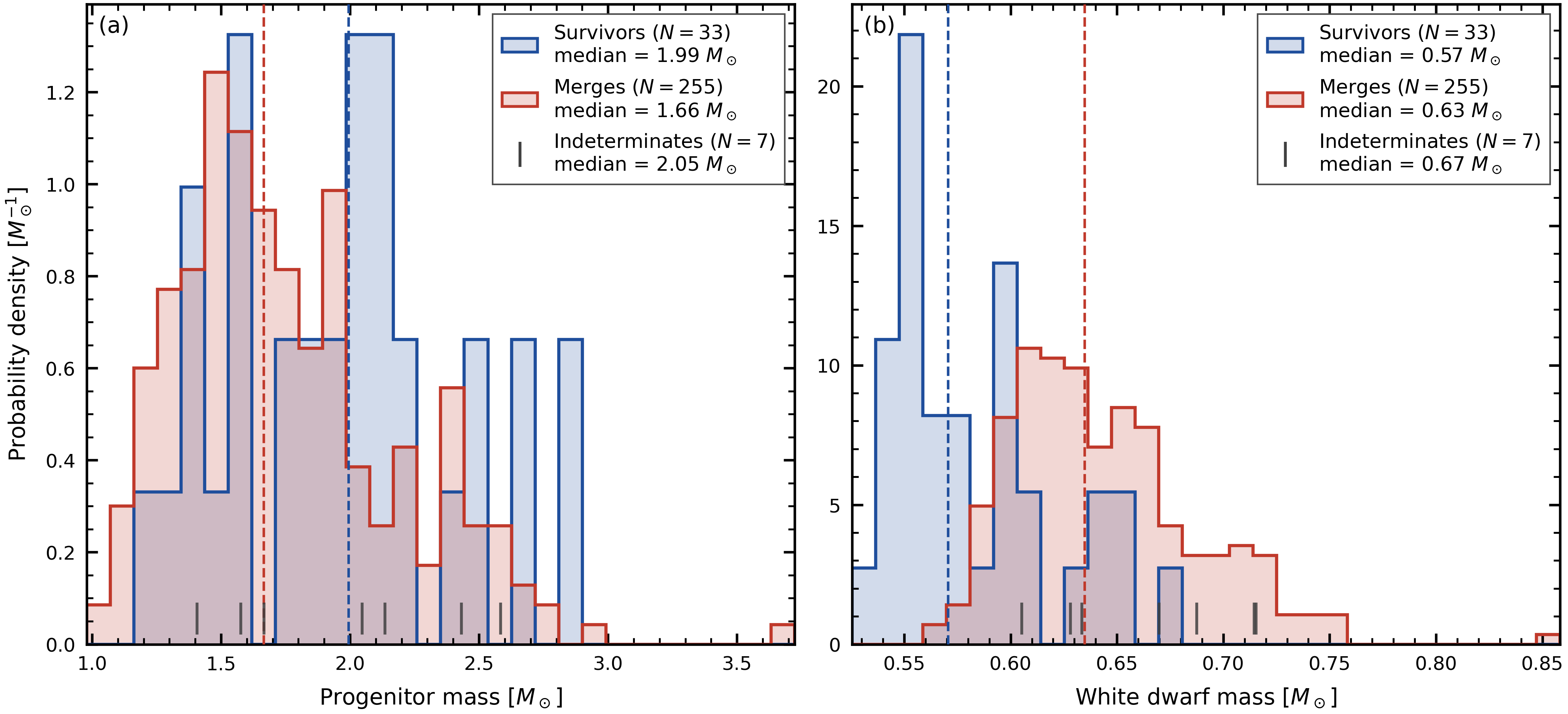}
    \caption{Normalized distributions of progenitor masses (a) and WD-mass estimates (b) for the $E_{\rm pot}+E_{\rm int}$ prescription with $\alpha_{\rm CE}=1$. Each histogram is normalized to unit area. Vertical dashed lines mark the median of each population. The seven indeterminate systems are shown individually as tick marks along the x-axis. Survivor WD masses are approximated by the core masses at CE onset. For merging and indeterminate systems, we used WD masses corresponding to single-star evolution. Survivors tend to originate from higher-mass progenitors but have lower estimated WD masses because CE ejection terminates subsequent core growth.}
    \label{fig:StellarMassDependence}
\end{figure*}

Figure~\ref{fig:InitialConditionsOutcomes} shows the outcomes of CE evolution as a function of the initial orbital separation, eccentricity, and companion mass for the different envelope-energy prescriptions. As expected, surviving companions preferentially originate from wider initial orbits and have relatively high masses. More massive companions release more orbital energy during inspiral, increasing their capability to eject the envelope before filling their Roche lobe. Similarly, companions initially located at larger orbital separations tend to be engulfed at later evolutionary stages, when the stellar envelope is generally less tightly bound. 
The outcome is therefore not only determined by companion mass or initial separation alone, but also depends sensitively on the stellar structure at the onset of CE evolution.

Figure~\ref{fig:StellarMassDependence} compares the progenitor and WD-mass distributions for the $E_{\rm pot}+E_{\rm int}$ prescription with $\alpha_{\rm CE}=1$. Survival is more frequent among higher-mass progenitors, where the survival fraction increases from $5.7^{+7.0}_{-3.3}$\% below $1.5\,M_\odot$ to $9.4^{+6.3}_{-4.0}$\% between $1.5$ and $2.0$\,$M_\odot$, and $20.7^{+12.1}_{-8.4}$\% between $2.0$ and $2.5\,M_\odot$. For progenitors with masses $\geq2.5\,M_\odot$, the fraction is $17.4^{+19.7}_{-10.4}\%$. The trend is therefore not strictly monotonic, and the highest-mass interval contains only $23$ CE systems, limiting the interpretation of its lower survival fraction. This association is consistent with the larger fraction of higher-mass progenitors reaching TP-AGB engulfment, when conditions are more favorable for survival. Panel (b) shows lower estimated WD masses for survivors, with a median of $0.57\,M_\odot$, compared with $0.63\,M_\odot$ for the merger group. Survivor WD masses are approximated by the core masses at CE onset, whereas the merger group uses WD masses from the parent population model, without accounting for how engulfment might alter subsequent stellar evolution. 

The six survivors obtained with the $E_{\rm pot}+E_{\rm therm}$ prescription and $\alpha_{\rm CE}=1$ 
span a broad range of initial progenitor masses, from $1.36$ to $2.58\,M_\odot$. A tendency towards smaller or larger envelope masses cannot be identified, potentially because of low-number statistics. 

\subsection{Properties of the surviving population}

\begin{figure}[t]
    \centering
    \includegraphics[width=0.9\linewidth]{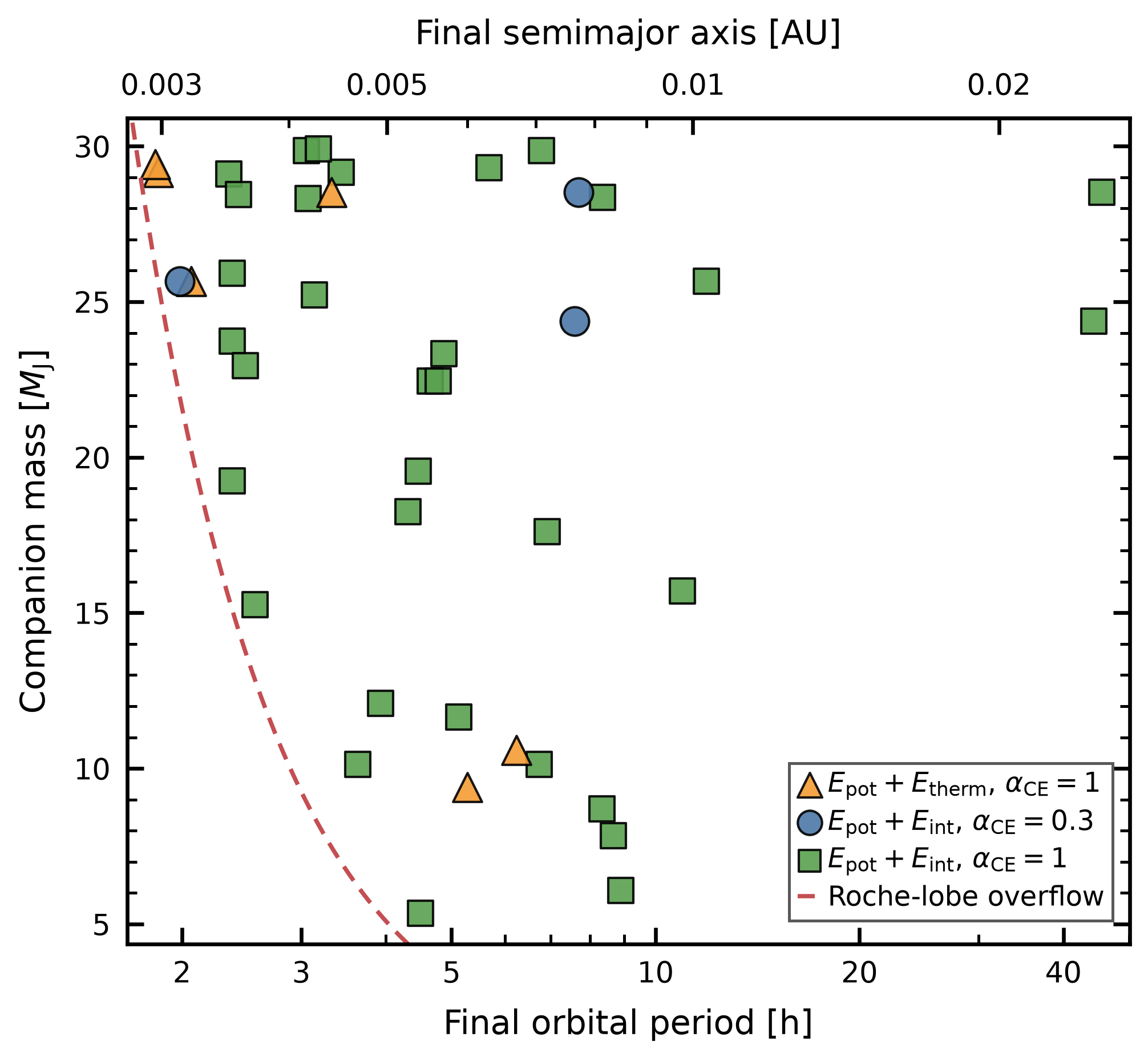}
    \caption{Companion masses and final orbital periods of post-CE survivors. Orange triangles represent the $E_{\rm pot}+E_{\rm therm}$ prescription with $\alpha_{\rm CE}=1$. Blue circles and green squares represent the $E_{\rm pot}+E_{\rm int}$ prescription with $\alpha_{\rm CE}=0.3$ and 1, respectively. The upper axis provides an approximate final semimajor axis using Kepler's third law with the median core and companion masses of all survivors, $\tilde{M}_{\rm c}=0.574\,M_\odot$ and $\tilde{M}_p=0.0216\,M_\odot$ (22.693 \Mjup), respectively. The red dashed curve marks the onset of companion RLOF, calculated using the Eggleton approximation and the median core mass quoted above. Most survivors have orbital periods of a few hours. However, including the full internal energy during CE evolution permits substantially wider final orbits.}
    \label{fig:FinalPeriod_vs_Mass}
\end{figure}

Figure~\ref{fig:FinalPeriod_vs_Mass} shows the masses and final orbital periods of the companions surviving CE evolution for the different envelope-energy prescriptions and CE efficiencies. As expected, the surviving systems have very compact orbits, with most orbital periods ranging from approximately 2 to 10 hours. However, a few systems reach periods of several tens of hours, illustrating the diversity of possible outcomes depending on the envelope structure at the onset of CE evolution.

Under the $E_{\rm pot}+E_{\rm therm}$ prescription, survival only occurs for $\alpha_{\rm CE}=1$, resulting in six systems with a median final orbital period of $2.69$\,h (Table~\ref{tab:main_results}). Interestingly, two of these six survivors have relatively low masses, despite their smaller orbital-energy reservoirs. This can be understood as the combined effect of the initial companion mass distribution and the strong dependence of envelope binding energy on the evolutionary stage at engulfment. Low-mass companions are more numerous in our adopted initial population, and although most cannot eject the stellar envelope, a small fraction enter CE evolution under particularly favorable conditions, close to the maximum stellar radius during a thermal pulse, when the envelope is only weakly bound. This illustrates that the post-CE mass distribution depends not only on the mass dependence of CE survival, but also on the initial companion distribution and the conditions at engulfment.

Adopting the $E_{\rm pot}+E_{\rm int}$ prescription significantly broadens the predicted population of surviving systems. For $\alpha_{\rm CE}=0.3$, three companions survive, all with masses in the brown-dwarf regime and final orbital periods of a few hours, with a median of $7.60$\,h. Increasing the efficiency to $\alpha_{\rm CE}=1$ produces a much larger and more diverse population, extending from companions with masses just above the adopted $5\,M_{\rm J}$ analysis threshold to brown dwarfs approaching $30\,M_{\rm J}$. Their final periods span approximately $2.3$--$45.6$\,h, with a median of $4.50$\,h.

The wide range of final orbital periods is a consequence of the strong variations in envelope binding energy during the TP-AGB, together with differences in the available orbital energy. Companions engulfed when the envelope is relatively tightly bound must spiral inward substantially before releasing sufficient orbital energy for ejection, resulting in very short final periods. In contrast, companions entering weakly bound envelopes can eject the envelope after a smaller reduction in orbital separation and consequently retain wider post-CE orbits. Thus, the final orbital period is not determined by companion mass alone, but also depends sensitively on the stellar structure at the onset of CE evolution.

Overall, our calculations predict that substellar companions surviving CE evolution should predominantly orbit WDs with periods of a few hours, although systems with periods approaching two days are possible under favorable conditions. The $E_{\rm pot}+E_{\rm int}$ prescription not only increases the predicted survival fraction but also extends the range of companion masses and orbital periods, allowing close WD companions in the planetary-mass regime above our adopted $5\,M_{\rm J}$ threshold.

\subsection{Predicted companions around WDs within 50\,pc}
\label{sect:50pcWDs}
To investigate how the predicted survival fractions translate into the observable WD population, we used a 50\,pc volume-limited sample taken from the \citet{GentileFusillo2021} WD catalog, selecting 1556 WDs with a WD probability above 95\% and masses within the range covered by our simulated population ($M_{\text{WD}}=0.5$--$0.85\,M_\odot$, assuming a pure-H composition). We estimated the expected number of surviving post-CE companions based on the occurrence fractions reported in Table~\ref{tab:main_results}, using $N_{\mathrm{postCE}} = N_{\mathrm{WD}}{f_{\rm CE}}$. These estimates therefore refer to the $M_p>5\,M_J$ companion population considered in our CE-survival analysis. Assuming that each WD independently has a probability $f_{\rm CE}$ of hosting a post-CE survivor, we estimate the probability of finding at least one companion $P(X\geq 1)$ using the binomial distribution, that is, $P(X\geq 1) = 1-\left(1-f_{\rm CE}\right)^{N_{\mathrm{WD}}}$. 

For the $E_{\rm pot}+E_{\rm therm}$ prescription with $\alpha_{\text{CE}}=1$, we obtain an expected number of 0.46 surviving companions within 50 pc, corresponding to a probability of approximately 37\% of finding at least one. For $E_{\rm pot}+E_{\rm int}$, the expected numbers are 0.23 and 2.53 for $\alpha_{\text{CE}}=0.3$ and 1, corresponding to probabilities of finding at least one companion within $50$~pc 
of approximately $21\%$ and $92\%$, respectively.

\subsection{Detectability of the surviving population}

We next assess whether the surviving companions predicted by our population synthesis could be detected with current observing facilities. We first considered transit searches under TESS-like observing conditions. For each surviving system, we approximated the WD radius using the zero-temperature mass--radius relation of \citet{Nauenberg1972}, ignoring finite-temperature corrections and adopting the post-CE core mass as the WD mass. We assumed a companion radius of \(R_{\rm p}=1\,R_{\rm J}\) and a circular post-CE orbit, giving a geometric transit probability of $p_{\rm transit}=(R_{\rm WD}+R_{\rm p})/a_{\rm f}$.

We generated synthetic light curves at cadences of 20 and 120\,s over a continuous 27.4\,d baseline, assuming Gaussian white noise with an RMS of 1\% in one hour. Transit recovery was evaluated through a blind Box Least Squares injection-recovery analysis following \citet{Faedi2011} and \citet{Robert2024}. Following the randomization procedure of \citet{Robert2024}, detection thresholds were calibrated separately for each cadence from 1000 random permutations of a Gaussian white-noise light curve, preserving the observing times. We adopted the empirical 99.9th percentile of the resulting maximum BLS-power distribution as the detection threshold. This empirical threshold is subject to finite Monte Carlo uncertainty. An injected transit was considered recovered when a candidate above this threshold matched the injected period, or an integer harmonic, within 2\%. The recovery sensitivity conditional on a transiting orientation, \(S\), was then combined with the geometric probability to obtain $p_{\rm detection}=p_{\rm transit}S$.

\begin{figure}[t]
    \centering
    \includegraphics[width=0.9\linewidth]{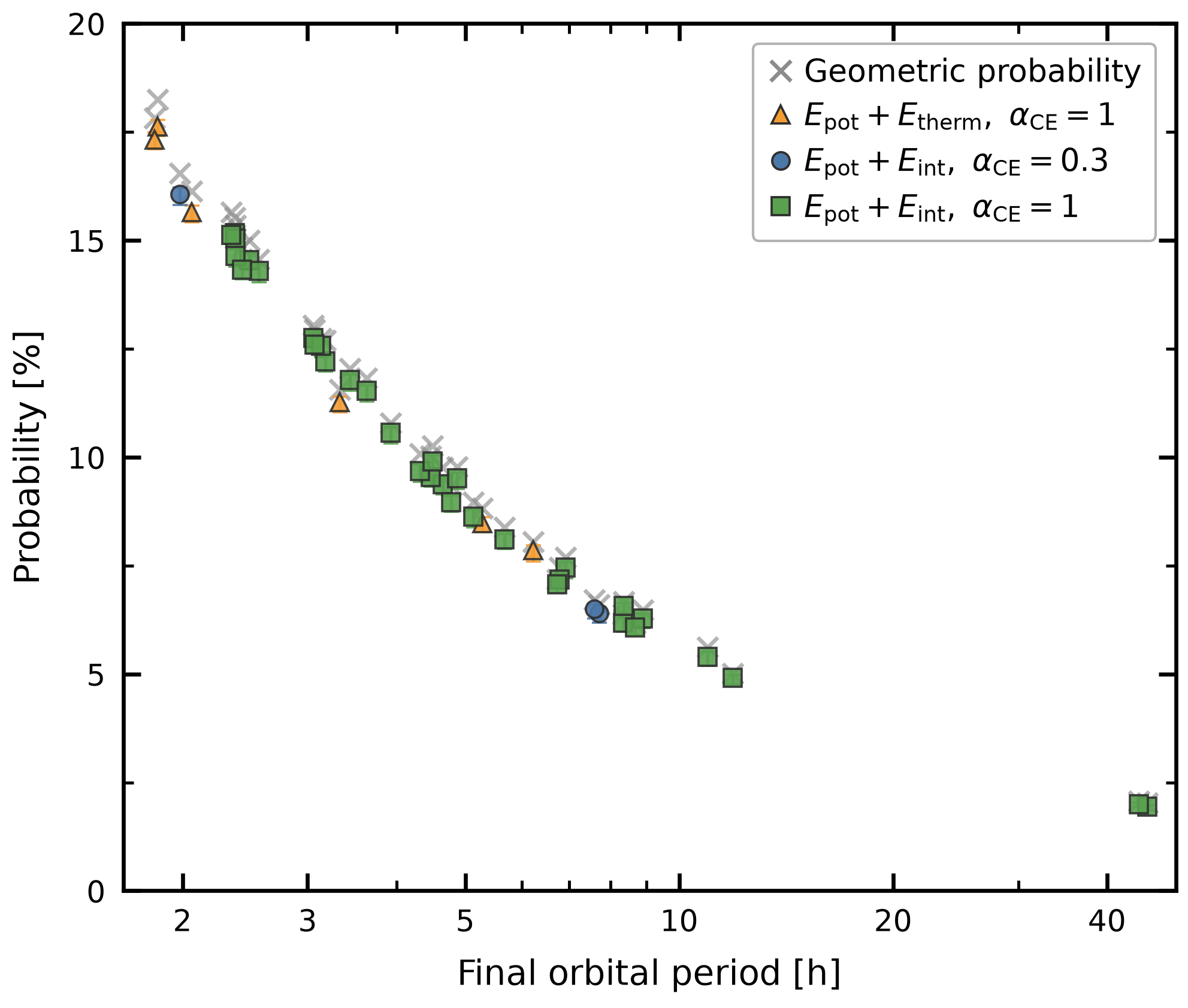}
    \caption{Predicted transit-detection probabilities for post-CE survivors at a TESS-like cadence of 120 s. Gray crosses represent the geometric probability, $p_{\rm transit}$, while symbols show $p_{\rm detection}$. Orange triangles represent the $E_{\rm pot}+E_{\rm therm}$ prescription with $\alpha_{\rm CE}=1$. Blue circles and green squares represent the $E_{\rm pot}+E_{\rm int}$ prescription with $\alpha_{\rm CE}=0.3$ and 1, respectively. Error bars show 95\% Wilson confidence intervals \citep{Wilson1927} on the injection-recovery sensitivity propagated to $p_{\rm detection}$. The simulations assume a continuous 27.4-d baseline and Gaussian white noise normalized to 1\% RMS in one hour. As expected, the detection probabilities lie close to the geometric transit probabilities, indicating that orbital orientation is the main limitation.}
    \label{fig:TESS}
\end{figure}

Figure\,\ref{fig:TESS} shows the resulting detection probabilities at 120\,s cadence. The mean geometric transit probabilities range from 9.95\% to 13.44\% (Table~\ref{tab:main_results}). The injected transits are recovered with mean sensitivities of approximately 97\% at both adopted cadences, resulting in mean detection probabilities that differ by at most 0.11 percentage points between 20 and 120\,s. Transit detectability is therefore limited primarily by the geometric probability of a favorable orbital inclination rather than by the observing cadence.

We also evaluated detectability through infrared excess with the James Webb Space Telescope (JWST). For the WDs, we adopted the cooling tracks of \citet{bedardetal2020} and the atmosphere models of \citet{Koester2010}. For the substellar companions, we used the Sonora Flame Skimmer atmospheric and evolutionary models \citep{Mang_2026}. Using the \texttt{species} package \citep{Stolker2020}, we modeled the spectral energy distributions of both components and computed synthetic photometry in four MIRI filters (for more details, see Mauch-Soriano et al., in prep.). Because of the very small final separations, the companion effective temperatures include both their intrinsic temperatures and the contribution from irradiation by the host WD, which increases the predicted infrared excess relative to isolated evolutionary models \citep{Poulsenetal2024}.

\begin{figure}[t]
    \centering
    \includegraphics[width=0.9\linewidth]{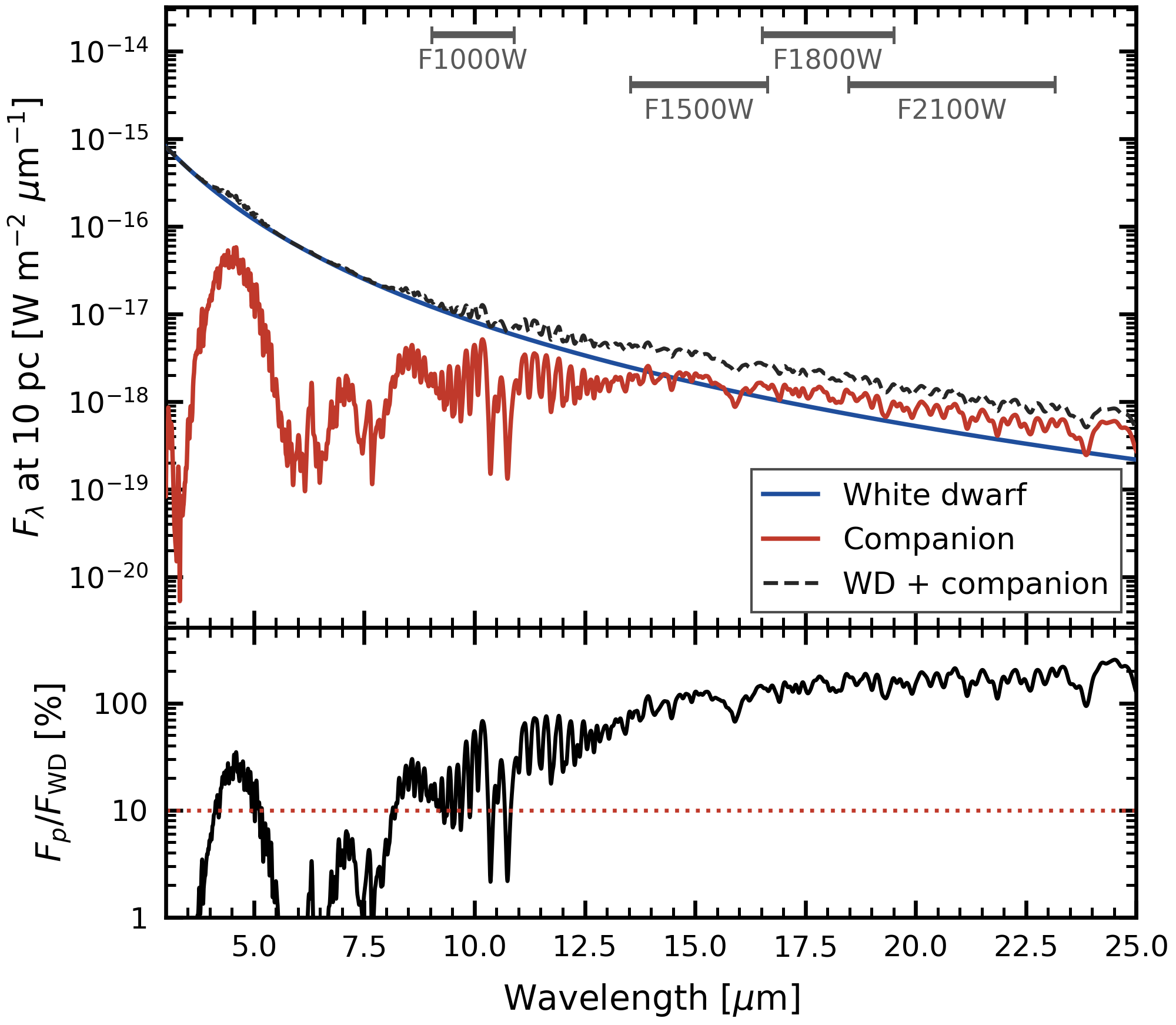}
    \caption{Top: Spectral energy distribution of one of our systems, showing the WD (blue), the companion (red), and their combined emission (black dashed curve). The horizontal bars indicate the half-power wavelength ranges of the MIRI filters, as reported in the JWST documentation. The WD parameters are \(M_{\mathrm{WD}}=0.61\,M_\odot\), \(T_{\mathrm{WD}}=5439\) K, and \(t_{\mathrm{cooling}}=3.74\) Gyr, while the companion is a planet with \(M_{\mathrm{p}}=9.4\,M_{\mathrm{J}}\), \(T_{\mathrm{p}}=386\) K, and \(a_{\mathrm{f}}=0.006\,{\mathrm{au}}\). Bottom: Infrared excess, expressed as the ratio of the planetary flux to the WD flux. Despite the companion's relatively modest mass, well within the planetary regime, its proximity to the WD results in a higher effective temperature, producing a significant infrared excess at wavelengths beyond \(8\,\mu\mathrm{m}\).} 
    \label{fig:ir_sed}
\end{figure}

We find that the infrared excess produced by all surviving companions would be detectable in F1500W $(15\,\mu{\rm m})$ at $>5\sigma$ out to 50 pc, assuming an integration time of $t=444\, s$ as adopted by the MEAD survey \citep{Poulsen_2026}. Moreover, over 95\% of the surviving companions would be detectable at $>3\sigma$ in F1000W ($10\,\mu{\rm m}$) under the same observing conditions, as well as in F1800W and F2100W ($18$ and $21\,\mu{\rm m}$, respectively), assuming integration times of $277$ and $710\,{\rm s}$, respectively, as adopted by the MEOW survey \citep{limbachetal24-1}. Figure~\ref{fig:ir_sed} illustrates the spectral energy distribution of one of the least massive surviving companions, showing that even a planetary-mass survivor can produce a substantial mid-infrared excess.

Since all predicted survivors would be detectable in F1500W under the adopted observing conditions, a complete survey of the selected WD sample within 50 pc would yield the detection expectations reported in Sect.~\ref{sect:50pcWDs}.

\section{Discussion}

We performed population synthesis calculations to investigate the survival and observable properties of post-CE substellar companions with WD progenitors. In what follows we briefly compare results with previous population-synthesis studies. 

Our calculations show that survival through CE evolution is rare and strongly depends on the evolutionary stage at engulfment. All surviving companions enter CE during the TP-AGB, when the stellar envelope can become sufficiently weakly bound to allow even planetary-mass companions to survive. The contribution of recombination energy substantially increases the predicted survival fraction, with occurrence rates reaching 0.16\% among WDs in our most optimistic model. The resulting companions typically have orbital periods of a few hours and should be detectable through mid-infrared excess with JWST. These predictions remain subject to uncertainties in the initial companion population, stellar tides, envelope ejection, and the adopted companion radius, as discussed in Appendix \ref{app:substellar_radius}. 

The survival of planetary companions during CE evolution has recently been investigated by \citet{gaiboretal26-1}, who concluded that surviving planets should be extremely rare, preferentially massive, and orbit very close to their host WDs. While our results broadly support the expectation that CE survival is restricted to a relatively small fraction of substellar companions, our approach differs in several important aspects and provides a more comprehensive prediction for the Galactic population of post-CE companions.

First, \citet{gaiboretal26-1} considered only two progenitor masses ($1$ and $2$~\Msun) at solar metallicity, whereas our population synthesis follows the evolution of progenitor stars spanning a continuous distribution of initial masses and metallicities. As in \citet{mauchsorianoetal26-1}, we generate a representative Galactic population using an initial mass function and an age--metallicity relation. This allows us to account for the dependence of stellar evolution, envelope structure, and CE survival on the properties of the progenitor stars when predicting the population of surviving companions.

Second, an important difference concerns the assumed initial population of substellar companions. \citet{gaiboretal26-1} adopted the planet mass and orbital separation distributions derived from the California Legacy Survey of predominantly solar-type main-sequence stars \citep{fultonetal21-1}, considering planetary masses up to $13$\,\Mjup. In contrast, we adopt the initial distributions and occurrence rates used by \citet{mauchsorianoetal26-1}, which are based on radial-velocity surveys of evolved giant stars \citep{wolthoff2022} and account for the observed dependence of giant-planet occurrence on stellar mass and metallicity. The companion-mass distribution, however, is based on observations of FGK main-sequence stars and is extrapolated into the brown-dwarf regime. Our companion population also extends to $30$\,\Mjup, including brown dwarfs. These differences are particularly relevant for CE evolution because survival strongly depends on both the companion mass and its initial orbital separation. Consequently, the adopted initial distributions affect not only the predicted number of surviving companions but also their mass and orbital-period distributions. Our approach therefore provides predictions based on a companion population specifically constrained for the progenitors of WDs, rather than extrapolating distributions derived primarily for solar-type main-sequence stars.

A further concern is the definition of the core-envelope boundary used to calculate the envelope binding energy. \citet{gaiboretal26-1} define the stellar core using an H mass-fraction threshold of $X_{\rm H}=0.1$, whereas we adopt a lower threshold of $X_{\rm H}=0.01$ \citep{scherbak+Fuller2023}. This places the core-envelope boundary deeper in the star, including more of the tightly bound envelope inner layers and therefore making envelope ejection more demanding in our calculations. This difference acts in the opposite direction to the inclusion of recombination energy, which reduces the net energy required to eject the envelope. 

In addition, we followed a mass-dependent increasing exponential overshooting at the convective core as suggested by \citet{claret&torres2017} for H-burning phases, and full-strength core-overshooting for all stellar masses during He-burning \citep{spruit2015,constantinoetal2015}. In contrast, \citet{gaiboretal26-1} only considered overshooting during the AGB at all convective boundaries and without any dependence on stellar mass. We tested the sensitivity of our calculations to the overshooting prescription and found that the implementation of the corrected treatment for overshooting has important effects on the stellar evolution, as it can modify the timing and properties of AGB thermal pulses, therefore affecting the conditions at CE onset and companion survival.

Another important difference is the treatment of the envelope binding energy itself and the conditions under which CE evolution begins. \citet{gaiboretal26-1} calculated the envelope binding energy from stellar evolution models but excluded recombination energy, exploring its possible contribution indirectly through enhanced CE ejection efficiencies, including values significantly exceeding unity. In contrast, we explicitly account for the contribution of recombination energy to the envelope binding energy. We find that this contribution can become particularly important during thermal pulses on the AGB, when the envelope is highly extended and only weakly bound. In some cases, the resulting envelope binding energy approaches zero from negative values, reducing the positive energy required for envelope ejection. If a substellar companion initiates CE evolution during such a weakly bound phase, the energy required to eject the envelope can be sufficiently small to allow the companion to survive without requiring an exceptionally high CE ejection efficiency. This identifies a potentially important survival channel that is not captured by calculations neglecting the contribution of recombination energy.

The survival probability therefore depends not only on the mass of the companion and the adopted CE efficiency, but also critically on the evolutionary stage at which engulfment occurs. In particular, the detailed evolution of the envelope structure during thermal pulses can create short intervals during which CE survival becomes considerably more likely. Since the occurrence and properties of these phases depend on the initial stellar mass and metallicity, accounting for the full distribution of WD progenitors is essential for predicting the population of surviving companions. We emphasize, however, that the contribution of recombination energy to envelope ejection remains uncertain, as not all of this energy may be available to unbind the envelope.

Finally, our study extends beyond predicting the intrinsic population of post-CE companions by investigating their detectability through infrared excess emission with JWST. Whereas \citet{gaiboretal26-1} focused primarily on the occurrence rates, orbital properties, and transit probabilities of surviving planets, we calculate the expected infrared contribution of substellar companions to the combined emission of WD--companion systems. This allows us to connect our population-synthesis predictions directly to observational searches and to identify the WD and companion properties for which surviving substellar objects are most likely to be detected.

Our calculations thus complement and extend the results of \citet{gaiboretal26-1} in two important ways. 
By accounting for a representative Galactic population of WD progenitors and substellar companions, we provide predictions for the occurrence rates and observable properties of post-CE systems. More importantly, by explicitly including recombination energy and resolving the evolutionary phases during which CE evolution is initiated, we identify conditions under which substellar companions can survive even when envelope ejection would otherwise appear energetically prohibitive.

With respect to the consistency of our findings with those of \citet{gaiboretal26-1}, we find general agreement: Post-common-envelope substellar companions should be scarce, massive, and have orbital periods of just a few hours. However, some quantitative estimates differ and are worth discussing. To properly compare our findings, we restrict our analysis to cases with $M_p<13$~\Mjup. 

For $E_{\rm pot}+E_{\rm therm}$ and $\alpha_{\rm CE}=1$, \citet{gaiboretal26-1} find no survivors around $1$\Msun\ progenitors and only 142 among $10^7$ companions around $2$\Msun\ progenitors when adopting variable tidal quality factors. In contrast, we find two planetary-mass survivors among 174 systems, both with $M_p\sim10$\Mjup\ around $\sim1.5$\Msun\ progenitors. However, differences in the progenitor masses and initial companion distributions prevent a direct quantitative comparison of the survival fractions.

A further difference concerns the final orbital separations. While \citet{gaiboretal26-1} find most survivors within $1.4\,a_{\rm disrupt}$, where their tidal-disruption separation $a_{\rm disrupt}$ is defined by the same Eggleton Roche-lobe criterion that we use to define $a_{\rm RLOF}$, and suggest this as a possible observational signature of CE evolution. Expressed in the same normalization, our thermal-energy model predicts two planetary-mass survivors at $1.47$ and $1.70\,a_{\rm RLOF}$. Including the full internal energy yields eight survivors spanning $1.10$--$1.94\,a_{\rm RLOF}$. Thus, $1.4\,a_{\rm RLOF}$ does not represent a universal upper boundary for CE survivors. Nevertheless, for our reference radius of $R_p=1\,R_{\rm J}$, all our planetary-mass survivors remain within $2\,a_{\rm RLOF}$, approximately the lower limit expected for high-eccentricity migration \citep{Ford_rasio_2006}, suggesting that the final orbital separation may still help distinguish the two formation channels.

\section{Conclusion}

We extended the population-synthesis framework of \citet{mauchsorianoetal26-1}, based on 20\,315 WD progenitors, to investigate the survival of 295 engulfed substellar companions with masses of $5$--$30$\,\Mjup. We considered two envelope-energy prescriptions and CE efficiencies of $\alpha_{\rm CE}=0.3$ and $1$, and predicted the occurrence rates, properties, and detectability of the resulting systems. 
Our main findings are:

1. CE survival is rare and strongly depends on the evolutionary stage at engulfment. None of the companions entering CE during the RGB or E-AGB survive. For $E_{\rm pot}+E_{\rm therm}$, no companions survive with $\alpha_{\rm CE}=0.3$, while six survive with $\alpha_{\rm CE}=1$. Including the full internal energy increases these numbers to three and 33, respectively. All survivors enter CE during the TP-AGB, when the stellar envelope can become weakly bound.

2. The predicted occurrence rates range from zero to $0.16\%$ among WDs, depending on the adopted CE prescription. Our most optimistic model predicts an average of 2.53 surviving companions among WDs within 50 pc, illustrating that CE evolution produces only a small population of close substellar companions.

3. Survival depends on the combination of envelope structure and available orbital energy, rather than companion mass alone. Even relatively low-mass companions can survive when engulfment occurs under favorable conditions during thermal pulses. For $E_{\rm pot}+E_{\rm int}$ and $\alpha_{\rm CE}=1$, survival among engulfed systems is more frequent in the higher-mass progenitor intervals, consistent with a larger fraction reaching TP-AGB contact. However, the sampled mass dependence is not strictly monotonic. Furthermore, including the full internal energy yields some formally non-negative envelope binding energies ($E_{\rm bind}\geq0$), for which the standard CE energy formalism cannot reliably predict the outcome.

4. Most surviving companions have final orbital periods of approximately $2$--$10\,{\rm h}$, although periods approaching two days are possible. The final separations reflect the envelope binding energy at engulfment, with weakly bound envelopes allowing survival at wider orbits.

5. Transit detection is primarily limited by the geometric probability of approximately $10\%$, whereas infrared-excess searches are less sensitive to orbital inclination. Under our adopted JWST observing conditions, all surviving companions would be detectable in F1500W at $>5\sigma$ out to 50 pc, and more than $95\%$ in F1000W, F1800W and F2100W at $>3\sigma$.

In summary, our population synthesis predicts that CE evolution provides a rare but potentially observable channel for forming close substellar companions to WDs, complementing the wider-orbit population investigated by \citet{mauchsorianoetal26-1}. The predicted occurrence depends critically on the evolutionary stage at engulfment and the uncertain contribution of internal energy to envelope ejection. Mid-infrared surveys of nearby WDs offer a promising test of these predictions.

\begin{acknowledgements}
AMS acknowledges support from Agencia Nacional de Investigación y Desarrollo (ANID) Becas/Doctorado Nacional 21263654. JABJ acknowledges support from ANID Becas/Doctorado Nacional 21263186. MRS and JABJ thank for support from the
research unit FOR2990 (eRO-STEP: Stellar endpoints with eROSITA) funded
by the Deutsche Forschungsgemeinschaft and the internal research project
PI\_LIR\_26\_07 at USM. FLV acknowledges support from ANID
FONDECYT Postdoctoral Grant No. 3250464.
\end{acknowledgements}

\bibliographystyle{aa}
\bibliography{references}

@article{mauchsorianoetal26-1,
  author = {{Mauch-Soriano}, Alex and {Schreiber}, Matthias R. and {Correa}, Diego and {Pinilla}, Julio and {Riveros-Jara}, Catalina and {Vivanco}, Javiera and {Ronco}, Maria Paula and {Belloni}, Diogo and {Lagos-Vilches}, Felipe and {Brandner}, Wolfgang},
  title = {{Predicted incidence of Jupiter-like planets around white dwarfs}},
  journal = {A\&A},
  year = {2026},
  volume = {707},
  pages = {A90},
  doi = {10.1051/0004-6361/202558367},
  archiveprefix = {arXiv},
  eprint = {2601.19531},
  primaryclass = {astro-ph.EP}
}

@misc{gaiboretal26-1,
  author = {{Gaibor}, Yadira S. and {Vanderburg}, Andrew and {Becker}, Juliette C. and {Soares-Furtado}, Melinda and {Rappaport}, Saul and {Daylan}, Tansu},
  title = {{Planets surviving common envelope evolution should be rare, massive, and orbit very close to their host star}},
  year = {2026},
  howpublished = {arXiv e-prints, arXiv:2610.27055},
  archiveprefix = {arXiv},
  eprint = {2610.27055},
  primaryclass = {astro-ph.EP},
  url = {https://arxiv.org/abs/2610.27055}
}

@article{ronco2020,
  author = {{Ronco}, Mar{\'i}a Paula and {Schreiber}, Matthias R. and {Giuppone}, Cristian A. and {Veras}, Dimitri and {Cuadra}, Jorge and {Guilera}, Octavio M.},
  title = {{How Jupiters Save or Destroy Inner Neptunes around Evolved Stars}},
  journal = {ApJ},
  year = {2020},
  volume = {898},
  number = {1},
  pages = {L23},
  doi = {10.3847/2041-8213/aba35f}
}

@article{jermyn2023,
  author = {{Jermyn}, Adam S. and {Bauer}, Evan B. and {Schwab}, Josiah and {Farmer}, R. and {Ball}, Warrick H. and {Bellinger}, Earl P. and {Dotter}, Aaron and {Joyce}, Meridith and {Marchant}, Pablo and {Mombarg}, Joey S. G. and {Wolf}, William M. and {Sunny Wong}, Tin Long and {Cinquegrana}, Giulia C. and {Farrell}, Eoin and {Smolec}, R. and {Thoul}, Anne and {Cantiello}, Matteo and {Herwig}, Falk and {Toloza}, Odette and {Bildsten}, Lars and {Townsend}, Richard H. D. and {Timmes}, F. X.},
  title = {{Modules for Experiments in Stellar Astrophysics (MESA): Time-dependent Convection, Energy Conservation, Automatic Differentiation, and Infrastructure}},
  journal = {ApJS},
  year = {2023},
  volume = {265},
  number = {1},
  pages = {15},
  doi = {10.3847/1538-4365/acae8d}
}

@article{LivioSoker1984,
  author = {{Livio}, M. and {Soker}, N.},
  title = {{Star--planet systems as possible progenitors of cataclysmic binaries}},
  journal = {MNRAS},
  year = {1984},
  volume = {208},
  number = {4},
  pages = {763--781},
  doi = {10.1093/mnras/208.4.763}
}

@article{NordhausSpiegel2013,
  author = {{Nordhaus}, J. and {Spiegel}, D. S.},
  title = {{On the orbits of low-mass companions to white dwarfs and the fates of the known exoplanets}},
  journal = {MNRAS},
  year = {2013},
  volume = {432},
  number = {1},
  pages = {500--505},
  doi = {10.1093/mnras/stt569}
}

@article{Karakas2014,
  author = {{Karakas}, Amanda I. and {Lattanzio}, John C.},
  title = {{The Dawes Review 2: Nucleosynthesis and stellar yields of low and intermediate-mass single stars}},
  journal = {PASA},
  year = {2014},
  volume = {31},
  pages = {e030},
  doi = {10.1017/pasa.2014.21}
}

@article{Parsons2025,
  author = {{Parsons}, S G and {Brown}, A J and {Casewell}, S L and {Littlefair}, S P and {van Roestel}, J and {Rebassa-Mansergas}, A and {Murillo-Ojeda}, R and {Hollands}, M A and {Zorotovic}, M and {Castro Segura}, N and {Dhillon}, V S and {Dyer}, M J and {Garbutt}, J A and {Green}, M J and {Jarvis}, D and {Kennedy}, M R and {Kerry}, P and {McCormac}, J and {Munday}, J and {Pelisoli}, I and {Pike}, E and {Sahman}, D I},
  title = {{Two almost planetary mass survivors of common envelope evolution}},
  journal = {MNRAS},
  year = {2025},
  volume = {537},
  number = {2},
  pages = {2112--2126},
  doi = {10.1093/mnras/staf166}
}

@article{williamsetal26,
  author = {{Williams}, Jamie T. and
            {G{\"a}nsicke}, Boris T. and
            {Stone}, Nicholas C. and
            {Koester}, Detlev and
            {Davies}, Benjamin D. R. and
            {Tong}, Christopher and
            {Wilson}, David J. and
            {Sahu}, Snehalata and
            {Swan}, Andrew and
            {Beatty}, Thomas G. and
            {Ram{\'i}rez}, Sergio H. and
            {Cunningham}, Tim and
            {Long}, Knox S.},
  title = {{Discovery of a second-generation planet candidate accreting onto a white dwarf}},
  journal = {Nature Astronomy},
  year = {2026},
  doi = {10.1038/s41550-026-02983-7},
  eprint = {2610.07161},
  archiveprefix = {arXiv},
  primaryclass = {astro-ph.EP}
}

@article{eggleton1983,
  author = {{Eggleton}, P. P.},
  title = {{Aproximations to the Radii of Roche Lobes}},
  journal = {ApJ},
  year = {1983},
  volume = {268},
  pages = {368},
  doi = {10.1086/160960}
}

@article{ibenlivio1993,
  author = {Iben, Icko, Jr. and Livio, Mario},
  title = {{Common envelopes in binary star evolution}},
  journal = {PASP},
  year = {1993},
  volume = {105},
  pages = {1373--1406},
  doi = {10.1086/133321}
}

@article{Belloni2024,
  author = {{Belloni}, Diogo and {Zorotovic}, Monica and {Schreiber}, Matthias R. and {Parsons}, Steven G. and {Moe}, Maxwell and {Garbutt}, James A.},
  title = {{Formation of long-period post-common envelope binaries. I. No extra energy is needed to explain oxygen-neon white dwarfs paired with AFGK-type main-sequence stars}},
  journal = {A\&A},
  year = {2024},
  volume = {686},
  pages = {A61},
  doi = {10.1051/0004-6361/202449235},
  eid = {A61}
}

@article{Zorotovic2010,
  author = {{Zorotovic}, M. and {Schreiber}, M. R. and {G{\"a}nsicke}, B. T. and {Nebot G{\'o}mez-Mor{\'a}n}, A.},
  title = {{Post-common-envelope binaries from SDSS. IX. Constraining the common-envelope efficiency}},
  journal = {A\&A},
  year = {2010},
  volume = {520},
  pages = {A86},
  doi = {10.1051/0004-6361/200913658}
}

@article{Toonen2012,
  author = {{Toonen}, S. and {Nelemans}, G. and {Portegies Zwart}, S.},
  title = {{Supernova Type Ia progenitors from merging double white dwarfs. Using a new population synthesis model}},
  journal = {A\&A},
  year = {2012},
  volume = {546},
  pages = {A70},
  doi = {10.1051/0004-6361/201218966}
}

@article{Zorotovic2022,
  author = {{Zorotovic}, Monica and {Schreiber}, MatthiasR},
  title = {{Close detached white dwarf + brown dwarf binaries: further evidence for low values of the common envelope efficiency}},
  journal = {MNRAS},
  year = {2022},
  volume = {513},
  number = {3},
  pages = {3587--3595},
  doi = {10.1093/mnras/stac1137}
}

@article{Webbink1984,
  author = {{Webbink}, R. F.},
  title = {{Double white dwarfs as progenitors of R Coronae Borealis stars and Type I supernovae}},
  journal = {ApJ},
  year = {1984},
  volume = {277},
  pages = {355},
  doi = {10.1086/161701}
}

@article{SiessLivio1999b,
  author = {{Siess}, L. and {Livio}, M.},
  title = {{The accretion of brown dwarfs and planets by giant stars -- II. Solar-mass stars on the red giant branch}},
  journal = {MNRAS},
  year = {1999},
  volume = {308},
  number = {4},
  pages = {1133--1149},
  doi = {10.1046/j.1365-8711.1999.02784.x}
}

@article{OConnor2023,
  author = {{O'Connor}, Christopher E. and {Bildsten}, Lars and {Cantiello}, Matteo and {Lai}, Dong},
  title = {{Giant Planet Engulfment by Evolved Giant Stars: Light Curves, Asteroseismology, and Survivability}},
  journal = {ApJ},
  year = {2023},
  volume = {950},
  number = {2},
  pages = {128},
  doi = {10.3847/1538-4357/acd2d4},
  eid = {128}
}

@article{JiaSpruit2018,
  author = {{Jia}, Shi and {Spruit}, H. C.},
  title = {{Disruption of a Planet Spiraling into its Host Star}},
  journal = {ApJ},
  year = {2018},
  volume = {864},
  number = {2},
  pages = {169},
  doi = {10.3847/1538-4357/aad77c},
  archiveprefix = {arXiv},
  eprint = {1808.00467},
  primaryclass = {astro-ph.EP}
}

@article{Yarza2023,
  author = {{Yarza}, Ricardo and {Razo-L{\'o}pez}, Naela B. and {Murguia-Berthier}, Ariadna and {Everson}, Rosa Wallace and {Antoni}, Andrea and {MacLeod}, Morgan and {Soares-Furtado}, Melinda and {Lee}, Dongwook and {Ramirez-Ruiz}, Enrico},
  title = {{Hydrodynamics and Survivability during Post-main-sequence Planetary Engulfment}},
  journal = {ApJ},
  year = {2023},
  volume = {954},
  number = {2},
  pages = {176},
  doi = {10.3847/1538-4357/acbdfc},
  archiveprefix = {arXiv},
  eprint = {2203.11227},
  primaryclass = {astro-ph.SR}
}

@article{Lau2025,
  author = {{Lau}, Mike Y. M. and {Cantiello}, Matteo and {Jermyn}, Adam S. and {MacLeod}, Morgan and {Mandel}, Ilya and {Price}, Daniel J.},
  title = {{Hot Jupiter engulfment by an early red giant in 3D hydrodynamics}},
  journal = {A\&A},
  year = {2025},
  volume = {694},
  pages = {A264},
  doi = {10.1051/0004-6361/202452081},
  eid = {A264}
}

@misc{Lau2026,
  author = {{Lau}, Mike Y. M. and {Andrassy}, Robert and {Leidi}, Giovanni and {Gagnier}, Damien and {Mor{\'a}n-Fraile}, Javier and {R{\"o}pke}, Friedrich K. and {Mandel}, Ilya},
  title = {{Continuous mass ablation of planets engulfed in stellar envelopes}},
  year = {2026},
  howpublished = {arXiv e-prints, arXiv:2605.11343},
  archiveprefix = {arXiv},
  eprint = {2605.11343},
  primaryclass = {astro-ph.SR}
}

@article{GonzalezBolivar2022,
  author = {{Gonz{\'a}lez-Bol{\'i}var}, Miguel and {De Marco}, Orsola and {Lau}, Mike Y M and {Hirai}, Ryosuke and {Price}, Daniel J},
  title = {{Common envelope binary interaction simulations between a thermally-pulsating AGB star and a low mass companion}},
  journal = {MNRAS},
  year = {2022},
  volume = {517},
  number = {3},
  pages = {3181--3199},
  doi = {10.1093/mnras/stac2301}
}

@article{Reichardt2020,
  author = {{Reichardt}, Thomas A and {De Marco}, Orsola and {Iaconi}, Roberto and {Chamandy}, Luke and {Price}, Daniel J},
  title = {{The impact of recombination energy on simulations of the common-envelope binary interaction}},
  journal = {MNRAS},
  year = {2020},
  volume = {494},
  number = {4},
  pages = {5333--5349},
  doi = {10.1093/mnras/staa937}
}

@article{Casewell2024,
  author = {{Casewell}, S L and {Burleigh}, M R and {Napiwotzki}, R and {Zorotovic}, M and {Bergeron}, P and {French}, J R and {Hermes}, J J and {Faedi}, F and {Lawrie}, K L},
  title = {{The evolutionary history of GD 1400AB, a white dwarf--brown dwarf binary}},
  journal = {MNRAS},
  year = {2024},
  volume = {535},
  number = {1},
  pages = {753--762},
  doi = {10.1093/mnras/stae2301}
}

@article{Robert2024,
  author = {{Robert}, Akshay and {Farihi}, Jay and {Van Eylen}, Vincent and {Aungwerojwit}, Amornrat and {G{\"a}nsicke}, Boris T. and {Redfield}, Seth and {Dhillon}, Vikram S. and {Marsh}, Thomas R. and {Swan}, Andrew},
  title = {{The frequency of transiting planetary systems around polluted white dwarfs}},
  journal = {MNRAS},
  year = {2024},
  volume = {533},
  number = {2},
  pages = {1756--1765},
  doi = {10.1093/mnras/stae1859},
  archiveprefix = {arXiv},
  eprint = {2407.21743},
  primaryclass = {astro-ph.EP}
}

@article{Faedi2011,
  author = {{Faedi}, F. and {West}, R. G. and {Burleigh}, M. R. and {Goad}, M. R. and {Hebb}, L.},
  title = {{Detection limits for close eclipsing and transiting substellar and planetary companions to white dwarfs in the WASP survey}},
  journal = {MNRAS},
  year = {2011},
  volume = {410},
  number = {2},
  pages = {899--911},
  doi = {10.1111/j.1365-2966.2010.17488.x}
}

@article{Nauenberg1972,
  author = {{Nauenberg}, Michael},
  title = {{Analytic Approximations to the Mass-Radius Relation and Energy of Zero-Temperature Stars}},
  journal = {ApJ},
  year = {1972},
  volume = {175},
  pages = {417},
  doi = {10.1086/151568}
}

@article{GentileFusillo2021,
  author = {{Gentile Fusillo}, N P and {Tremblay}, P-E and {Cukanovaite}, E and {Vorontseva}, A and {Lallement}, R and {Hollands}, M and {G{\"a}nsicke}, B T and {Burdge}, K B and {McCleery}, J and {Jordan}, S},
  title = {{A catalogue of white dwarfs in Gaia EDR3}},
  journal = {MNRAS},
  year = {2021},
  volume = {508},
  number = {3},
  pages = {3877--3896},
  doi = {10.1093/mnras/stab2672}
}

@article{Clayton_2017,
  author = {{Clayton}, Matthew and {Podsiadlowski}, Philipp and {Ivanova}, Natasha and {Justham}, Stephen},
  title = {{Episodic mass ejections from common-envelope objects}},
  journal = {MNRAS},
  year = {2017},
  volume = {470},
  number = {2},
  pages = {1788--1808},
  doi = {10.1093/mnras/stx1290},
  url = {http://dx.doi.org/10.1093/mnras/stx1290}
}

@article{Wilson1927,
  author = {{Wilson}, Edwin B.},
  title = {{Probable Inference, the Law of Succession, and Statistical Inference}},
  journal = {J. Am. Stat. Assoc.},
  year = {1927},
  volume = {22},
  number = {158},
  pages = {209--212},
  doi = {10.1080/01621459.1927.10502953}
}

@article{adamsetal21-1,
  author = {{Adams}, Fred C. and {Meyer}, Michael R. and {Adams}, Arthur D.},
  title = {{A Theoretical Framework for the Mass Distribution of Gas Giant Planets Forming through the Core Accretion Paradigm}},
  journal = {ApJ},
  year = {2021},
  volume = {909},
  number = {1},
  pages = {1},
  doi = {10.3847/1538-4357/abdd2b},
  archiveprefix = {arXiv},
  eprint = {2101.06714},
  primaryclass = {astro-ph.EP},
  eid = {1}
}

@article{andryushin+popov21-1,
  author = {{Andryushin}, A. S. and {Popov}, S. B.},
  title = {{Exoplanet Population Synthesis with Account for Orbit Variation Due to Stellar Evolution}},
  journal = {Astron. Rep.},
  year = {2021},
  volume = {65},
  number = {4},
  pages = {246--268},
  doi = {10.1134/S1063772921040016},
  archiveprefix = {arXiv},
  eprint = {2309.12635},
  primaryclass = {astro-ph.EP}
}

@article{bedardetal2020,
  author = {{B{\'e}dard}, A. and {Bergeron}, P. and {Brassard}, P. and {Fontaine}, G.},
  title = {{On the Spectral Evolution of Hot White Dwarf Stars. I. A Detailed Model Atmosphere Analysis of Hot White Dwarfs from SDSS DR12}},
  journal = {ApJ},
  year = {2020},
  volume = {901},
  number = {2},
  pages = {93},
  doi = {10.3847/1538-4357/abafbe},
  archiveprefix = {arXiv},
  eprint = {2008.07469},
  primaryclass = {astro-ph.SR},
  eid = {93}
}

@article{Baraffe2003,
  author  = {Baraffe, I. and Chabrier, G. and Barman, T. S. and
             Allard, F. and Hauschildt, P. H.},
  title   = {Evolutionary models for cool brown dwarfs and extrasolar
             giant planets. The case of {HD 209458}},
  journal = {A\&A},
  year    = {2003},
  volume  = {402},
  pages   = {701--712},
  doi     = {10.1051/0004-6361:20030252}
}

@article{cassanetal12-1,
  author = {{Cassan}, A. and {Kubas}, D. and {Beaulieu}, J.-P. and {Dominik}, M. and {Horne}, K. and {Greenhill}, J. and {Wambsganss}, J. and {Menzies}, J. and {Williams}, A. and {J{\o}rgensen}, U. G. and {Udalski}, A. and {Bennett}, D. P. and {Albrow}, M. D. and {Batista}, V. and {Brillant}, S. and {Caldwell}, J. A. R. and {Cole}, A. and {Coutures}, Ch. and {Cook}, K. H. and {Dieters}, S. and {Prester}, D. Dominis and {Donatowicz}, J. and {Fouqu{\'e}}, P. and {Hill}, K. and {Kains}, N. and {Kane}, S. and {Marquette}, J.-B. and {Martin}, R. and {Pollard}, K. R. and {Sahu}, K. C. and {Vinter}, C. and {Warren}, D. and {Watson}, B. and {Zub}, M. and {Sumi}, T. and {Szyma{\'n}ski}, M. K. and {Kubiak}, M. and {Poleski}, R. and {Soszynski}, I. and {Ulaczyk}, K. and {Pietrzy{\'n}ski}, G. and {Wyrzykowski}, {\L}.},
  title = {{One or more bound planets per Milky Way star from microlensing observations}},
  journal = {Nature},
  year = {2012},
  volume = {481},
  number = {7380},
  pages = {167--169},
  doi = {10.1038/nature10684},
  archiveprefix = {arXiv},
  eprint = {1202.0903},
  primaryclass = {astro-ph.EP}
}

@article{Chamandy-21,
  author = {{Chamandy}, Luke and {Blackman}, Eric G and {Nordhaus}, Jason and {Wilson}, Emily},
  title = {{Successive common envelope events from multiple planets}},
  journal = {MNRAS},
  year = {2021},
  volume = {502},
  number = {1},
  pages = {L110--L114},
  doi = {10.1093/mnrasl/slab017},
  archiveprefix = {arXiv},
  eprint = {2011.11106},
  primaryclass = {astro-ph.SR}
}

@article{cummingetal08-1,
  author = {{Cumming}, Andrew and {Butler}, R. Paul and {Marcy}, Geoffrey W. and {Vogt}, Steven S. and {Wright}, Jason T. and {Fischer}, Debra A.},
  title = {{The Keck Planet Search: Detectability and the Minimum Mass and Orbital Period Distribution of Extrasolar Planets}},
  journal = {PASP},
  year = {2008},
  volume = {120},
  number = {867},
  pages = {531--554},
  doi = {10.1086/588487},
  archiveprefix = {arXiv},
  eprint = {0803.3357},
  primaryclass = {astro-ph}
}

@article{Esseldeurs2026,
  author = {{Esseldeurs}, M. and {Mathis}, S. and {Decin}, L.},
  title = {{The fate of Earth during the Sun's giant phases: New constraints from ab initio tidal modelling and AGB mass loss}},
  journal = {A\&A},
  year = {2026},
  volume = {710},
  pages = {L26},
  doi = {10.1051/0004-6361/202660576}
}

@article{Ford_rasio_2006,
  author = {{Ford}, Eric B. and {Rasio}, Frederic A.},
  title = {{On the Relation between Hot Jupiters and the Roche Limit}},
  journal = {ApJ},
  year = {2006},
  volume = {638},
  number = {1},
  pages = {L45--L48},
  doi = {10.1086/500734},
  archiveprefix = {arXiv},
  eprint = {astro-ph/0512632},
  primaryclass = {astro-ph}
}

@article{fultonetal21-1,
  author = {{Fulton}, Benjamin J. and {Rosenthal}, Lee J. and {Hirsch}, Lea A. and {Isaacson}, Howard and {Howard}, Andrew W. and {Dedrick}, Cayla M. and {Sherstyuk}, Ilya A. and {Blunt}, Sarah C. and {Petigura}, Erik A. and {Knutson}, Heather A. and {Behmard}, Aida and {Chontos}, Ashley and {Crepp}, Justin R. and {Crossfield}, Ian J. M. and {Dalba}, Paul A. and {Fischer}, Debra A. and {Henry}, Gregory W. and {Kane}, Stephen R. and {Kosiarek}, Molly and {Marcy}, Geoffrey W. and {Rubenzahl}, Ryan A. and {Weiss}, Lauren M. and {Wright}, Jason T.},
  title = {{California Legacy Survey. II. Occurrence of Giant Planets beyond the Ice Line}},
  journal = {ApJS},
  year = {2021},
  volume = {255},
  number = {1},
  pages = {14},
  doi = {10.3847/1538-4365/abfcc1},
  archiveprefix = {arXiv},
  eprint = {2105.11584},
  primaryclass = {astro-ph.EP},
  eid = {14}
}

@article{gaensickeetal19-1,
  author = {{G{\"a}nsicke}, Boris T. and {Schreiber}, Matthias R. and {Toloza}, Odette and {Fusillo}, Nicola P. Gentile and {Koester}, Detlev and {Manser}, Christopher J.},
  title = {{Accretion of a giant planet onto a white dwarf star}},
  journal = {Nature},
  year = {2019},
  volume = {576},
  number = {7785},
  pages = {61--64},
  doi = {10.1038/s41586-019-1789-8},
  archiveprefix = {arXiv},
  eprint = {1912.01611},
  primaryclass = {astro-ph.EP}
}

@article{Koester2010,
  author = {{Koester}, D.},
  title = {{White dwarf spectra and atmosphere models}},
  journal = {Mem. Soc. Astron. Ital.},
  year = {2010},
  volume = {81},
  pages = {921--931}
}

@article{lagosetal21-1,
  author = {{Lagos}, F and {Schreiber}, M R and {Zorotovic}, M and {G{\"a}nsicke}, B T and {Ronco}, M P and {Hamers}, Adrian S},
  title = {{{WD} 1856 b: a close giant planet around a white dwarf that could have survived a common envelope phase}},
  journal = {MNRAS},
  year = {2021},
  volume = {501},
  number = {1},
  pages = {676--682},
  doi = {10.1093/mnras/staa3703},
  archiveprefix = {arXiv},
  eprint = {2010.09747},
  primaryclass = {astro-ph.EP}
}

@article{limbachetal24-1,
  author = {{Limbach}, Mary Anne and {Vanderburg}, Andrew and {Venner}, Alexander and {Blouin}, Simon and {Stevenson}, Kevin B. and {MacDonald}, Ryan J. and {Jenkins}, Sydney and {Bowens-Rubin}, Rachel and {Soares-Furtado}, Melinda and {Morley}, Caroline and {Janson}, Markus and {Debes}, John and {Xu}, Siyi and {Kleisioti}, Evangelia and {Kenworthy}, Matthew and {Butler}, Paul and {Crane}, Jeffrey D. and {Osip}, Dave and {Shectman}, Stephen and {Teske}, Johanna},
  title = {{The MIRI Exoplanets Orbiting White dwarfs (MEOW) Survey: Mid-infrared Excess Reveals a Giant Planet Candidate around a Nearby White Dwarf}},
  journal = {ApJ},
  year = {2024},
  volume = {973},
  number = {1},
  pages = {L11},
  doi = {10.3847/2041-8213/ad74ed},
  archiveprefix = {arXiv},
  eprint = {2408.16813},
  primaryclass = {astro-ph.EP},
  eid = {L11}
}

@article{macdonald2026aerosols,
  author = {{MacDonald}, Ryan J. and {O'Connor}, Christopher E. and {Boehm}, Victoria A. and {May}, E. M. and {Sing}, David K. and {Mullens}, Elijah and {Mayorga}, L. C. and {Foote}, Trevor O. and {Blouin}, Simon and {Pearce}, Logan A. and {Lewis}, Nikole K. and {Valenti}, Jeff and {Batalha}, Natasha E. and {Lally}, Maura and {Lothringer}, Joshua D. and {Marley}, Mark S. and {Mishra}, Ishan and {Mullally}, Susan E.},
  title = {{Aerosols and hydrocarbons in the atmosphere of a white dwarf planet}},
  journal = {Nature},
  year = {2026},
  volume = {655},
  number = {8121},
  pages = {76--80},
  doi = {10.1038/s41586-026-10514-7}
}

@ARTICLE{2026A&A...713A.228V,
       author = {{Villarreal D'Angelo}, C. and {Ronco}, M.~P. and {Schreiber}, M.~R. and {Toloza}, O. and {Esquivel}, A. and {G{\"a}nsicke}, B.~T.},
        title = "{Planetary atmospheric escape and disk formation around WDJ0914+1914}",
      journal = {\aap},
         year = 2026,
        month = sep,
       volume = {713},
          eid = {A228},
        pages = {A228},
          doi = {10.1051/0004-6361/202557082},
archivePrefix = {arXiv},
       eprint = {2608.00318},
 primaryClass = {astro-ph.EP},
       adsurl = {https://ui.adsabs.harvard.edu/abs/2026A&A...713A.228V}
}

@article{maldonadoetal21-1,
  author = {{Maldonado}, R F and {Villaver}, E and {Mustill}, A J and {Ch{\'a}vez}, M and {Bertone}, E},
  title = {{Do instabilities in high-multiplicity systems explain the existence of close-in white dwarf planets?}},
  journal = {MNRAS},
  year = {2021},
  volume = {501},
  number = {1},
  pages = {L43--L48},
  doi = {10.1093/mnrasl/slaa193},
  archiveprefix = {arXiv},
  eprint = {2010.11403},
  primaryclass = {astro-ph.EP}
}

@article{Mang_2026,
  author = {{Mang}, James and {Chachan}, Yayaati and {Morley}, Caroline V. and {Batalha}, Natasha E. and {Wogan}, Nicholas F. and {Mukherjee}, Sagnick and {Fortney}, Jonathan J. and {Marley}, Mark S. and {Visscher}, Channon and {Gharib-Nezhad}, Ehsan},
  title = {{The Sonora Substellar Atmosphere Models. VII. Flame Skimmer: Cloud-free Atmospheric and Evolutionary Models for the Coldest Substellar Objects}},
  journal = {ApJ},
  year = {2026},
  volume = {1008},
  number = {2},
  pages = {254},
  doi = {10.3847/1538-4357/ae9609}
}

@misc{mayoretal11-1,
  author = {{Mayor}, M. and {Marmier}, M. and {Lovis}, C. and {Udry}, S. and {S{\'e}gransan}, D. and {Pepe}, F. and {Benz}, W. and {Bertaux}, J. -L. and {Bouchy}, F. and {Dumusque}, X. and {Lo Curto}, G. and {Mordasini}, C. and {Queloz}, D. and {Santos}, N. C.},
  title = {{The HARPS search for southern extra-solar planets XXXIV. Occurrence, mass distribution and orbital properties of super-Earths and Neptune-mass planets}},
  year = {2011},
  howpublished = {arXiv e-prints, arXiv:1109.2497},
  doi = {10.48550/arXiv.1109.2497},
  archiveprefix = {arXiv},
  eprint = {1109.2497},
  primaryclass = {astro-ph.EP}
}

@article{munoz+petrovic20-1,
  author = {{Mu{\~n}oz}, Diego J. and {Petrovich}, Cristobal},
  title = {{Kozai Migration Naturally Explains the White Dwarf Planet WD1856 b}},
  journal = {ApJ},
  year = {2020},
  volume = {904},
  number = {1},
  pages = {L3},
  doi = {10.3847/2041-8213/abc564},
  archiveprefix = {arXiv},
  eprint = {2010.04724},
  primaryclass = {astro-ph.EP},
  eid = {L3}
}

@article{Mustill_2012,
  author = {{Mustill}, Alexander J. and {Villaver}, Eva},
  title = {{FORETELLINGS OF RAGNAR{\"O}K: WORLD-ENGULFING ASYMPTOTIC GIANTS AND THE INHERITANCE OF WHITE DWARFS}},
  journal = {ApJ},
  year = {2012},
  volume = {761},
  number = {2},
  pages = {121},
  doi = {10.1088/0004-637X/761/2/121},
  url = {https://dx.doi.org/10.1088/0004-637X/761/2/121}
}

@article{Nelemans-98,
  author = {{Nelemans}, G. and {Tauris}, T. M.},
  title = {{Formation of undermassive single white dwarfs and the influence of planets on late stellar evolution}},
  journal = {A\&A},
  year = {1998},
  volume = {335},
  pages = {L85--L88},
  doi = {10.48550/arXiv.astro-ph/9806011},
  archiveprefix = {arXiv},
  eprint = {astro-ph/9806011},
  primaryclass = {astro-ph}
}

@article{Poulsen_2026,
  author = {{Poulsen}, Sabrina and {Debes}, John and {Messier}, Ashley and {Le Bourdais}, {\'E}rika and {Melis}, Carl and {Cracraft}, Misty and {Boucher}, Samuel and {Kilic}, Mukremin and {Kenyon}, Scott and {Wyatt}, Mark C. and {Redfield}, Seth and {Dufour}, Patrick and {Albert}, Loic and {Mullally}, Susan E. and {Reach}, William T. and {Mullally}, Fergal and {Golimowski}, David A.},
  title = {{The MIRI Excesses Around Degenerates (MEAD) Survey. II. A Probable Planet Detected via IR Excess around WD 0644+025}},
  journal = {AJ},
  year = {2026},
  volume = {171},
  number = {3},
  pages = {183},
  doi = {10.3847/1538-3881/ae3c0e}
}

@article{Poulsenetal2024,
  author = {{Poulsen}, Sabrina and {Debes}, John and {Cracraft}, Misty and {Mullally}, Susan E. and {Reach}, William T. and {Kilic}, Mukremin and {Mullally}, Fergal and {Albert}, Loic and {Thibault}, Katherine and {Hermes}, J. J. and {Barclay}, Thomas and {Quintana}, Elisa V.},
  title = {{A MIRI Search for Planets and Dust around WD 2149+021}},
  journal = {AJ},
  year = {2024},
  volume = {167},
  number = {6},
  pages = {257},
  doi = {10.3847/1538-3881/ad374c},
  archiveprefix = {arXiv},
  eprint = {2311.14165},
  primaryclass = {astro-ph.EP},
  eid = {257}
}

@article{rasioetal96-1,
  author = {{Rasio}, F. A. and {Tout}, C. A. and {Lubow}, S. H. and {Livio}, M.},
  title = {{Tidal Decay of Close Planetary Orbits}},
  journal = {ApJ},
  year = {1996},
  volume = {470},
  pages = {1187},
  doi = {10.1086/177941},
  archiveprefix = {arXiv},
  eprint = {astro-ph/9605059},
  primaryclass = {astro-ph}
}

@article{Stolker2020,
  author = {{Stolker}, T. and {Quanz}, S. P. and {Todorov}, K. O. and {K{\"u}hn}, J. and {Molli{\`e}re}, P. and {Meyer}, M. R. and {Currie}, T. and {Daemgen}, S. and {Lavie}, B.},
  title = {{MIRACLES: atmospheric characterization of directly imaged planets and substellar companions at 4-5 {\ensuremath{\mu}}m. I. Photometric analysis of {\ensuremath{\beta}} Pic b, HIP 65426 b, PZ Tel B, and HD 206893 B}},
  journal = {A\&A},
  year = {2020},
  volume = {635},
  pages = {A182},
  doi = {10.1051/0004-6361/201937159},
  archiveprefix = {arXiv},
  eprint = {1912.13316},
  primaryclass = {astro-ph.EP},
  eid = {A182}
}

@article{vanderburgetal20-1,
  author = {{Vanderburg}, Andrew and {Rappaport}, Saul A. and {Xu}, Siyi and {Crossfield}, Ian J. M. and {Becker}, Juliette C. and {Gary}, Bruce and {Murgas}, Felipe and {Blouin}, Simon and {Kaye}, Thomas G. and {Palle}, Enric and {Melis}, Carl and {Morris}, Brett M. and {Kreidberg}, Laura and {Gorjian}, Varoujan and {Morley}, Caroline V. and {Mann}, Andrew W. and {Parviainen}, Hannu and {Pearce}, Logan A. and {Newton}, Elisabeth R. and {Carrillo}, Andreia and {Zuckerman}, Ben and {Nelson}, Lorne and {Zeimann}, Greg and {Brown}, Warren R. and {Tronsgaard}, Ren{\'e} and {Klein}, Beth and {Ricker}, George R. and {Vanderspek}, Roland K. and {Latham}, David W. and {Seager}, Sara and {Winn}, Joshua N. and {Jenkins}, Jon M. and {Adams}, Fred C. and {Benneke}, Bj{\"o}rn and {Berardo}, David and {Buchhave}, Lars A. and {Caldwell}, Douglas A. and {Christiansen}, Jessie L. and {Collins}, Karen A. and {Col{\'o}n}, Knicole D. and {Daylan}, Tansu and {Doty}, John and {Doyle}, Alexandra E. and {Dragomir}, Diana and {Dressing}, Courtney and {Dufour}, Patrick and {Fukui}, Akihiko and {Glidden}, Ana and {Guerrero}, Natalia M. and {Guo}, Xueying and {Heng}, Kevin and {Henriksen}, Andreea I. and {Huang}, Chelsea X. and {Kaltenegger}, Lisa and {Kane}, Stephen R. and {Lewis}, John A. and {Lissauer}, Jack J. and {Morales}, Farisa and {Narita}, Norio and {Pepper}, Joshua and {Rose}, Mark E. and {Smith}, Jeffrey C. and {Stassun}, Keivan G. and {Yu}, Liang},
  title = {{A giant planet candidate transiting a white dwarf}},
  journal = {Nature},
  year = {2020},
  volume = {585},
  number = {7825},
  pages = {363--367},
  doi = {10.1038/s41586-020-2713-y},
  archiveprefix = {arXiv},
  eprint = {2009.07282},
  primaryclass = {astro-ph.EP}
}

@article{veras16-1,
  author = {{Veras}, Dimitri},
  title = {{Post-main-sequence planetary system evolution}},
  journal = {R. Soc. Open Sci.},
  year = {2016},
  volume = {3},
  pages = {150571},
  doi = {10.1098/rsos.150571},
  archiveprefix = {arXiv},
  eprint = {1601.05419},
  primaryclass = {astro-ph.EP},
  eid = {150571}
}

@article{villaver+livio07-1,
  author = {{Villaver}, Eva and {Livio}, Mario},
  title = {{Can Planets Survive Stellar Evolution?}},
  journal = {ApJ},
  year = {2007},
  volume = {661},
  number = {2},
  pages = {1192--1201},
  doi = {10.1086/516746},
  archiveprefix = {arXiv},
  eprint = {astro-ph/0702724},
  primaryclass = {astro-ph}
}

@article{Villaver_2009,
  author = {{Villaver}, Eva and {Livio}, Mario},
  title = {{THE ORBITAL EVOLUTION OF GAS GIANT PLANETS AROUND GIANT STARS}},
  journal = {ApJ},
  year = {2009},
  volume = {705},
  number = {1},
  pages = {L81--L85},
  doi = {10.1088/0004-637X/705/1/L81},
  url = {https://dx.doi.org/10.1088/0004-637X/705/1/L81}
}

@article{wolthoff2022,
  author = {{Wolthoff}, Vera and {Reffert}, Sabine and {Quirrenbach}, Andreas and {Jones}, Mat{\'i}as I. and {Wittenmyer}, Robert A. and {Jenkins}, James S.},
  title = {{Precise Radial Velocities of Giant Stars XVI: Planet Occurrence Rates from the Combined Analysis of the Lick, EXPRESS, and PPPS Giant Star Surveys}},
  journal = {A\&A},
  year = {2022},
  volume = {661},
  pages = {A63},
  doi = {10.1051/0004-6361/202142501}
}

@article{claret&torres2017,
  author = {{Claret}, Antonio and {Torres}, Guillermo},
  title = {{The Dependence of Convective Core Overshooting on Stellar Mass: A Semi-empirical Determination Using the Diffusive Approach with Two Different Element Mixtures}},
  journal = {ApJ},
  year = {2017},
  volume = {849},
  number = {1},
  pages = {18},
  doi = {10.3847/1538-4357/aa8770}
}

@article{constantinoetal2015,
  author = {{Constantino}, Thomas and {Campbell}, Simon W. and {Christensen-Dalsgaard}, J{\o}rgen and {Lattanzio}, John C. and {Stello}, Dennis},
  title = {{The treatment of mixing in core helium burning models -- I. Implications for asteroseismology}},
  journal = {MNRAS},
  year = {2015},
  volume = {452},
  number = {1},
  pages = {123--145},
  doi = {10.1093/mnras/stv1264}
}

@article{spruit2015,
  author = {{Spruit}, H. C.},
  title = {{The growth of helium-burning cores}},
  journal = {A\&A},
  year = {2015},
  volume = {582},
  pages = {L2},
  doi = {10.1051/0004-6361/201527171}
}

@article{scherbak+Fuller2023,
  title={White dwarf binaries suggest a common envelope efficiency $\alpha$~ 1/3},
  author={Scherbak, Peter and Fuller, Jim},
  journal={Monthly Notices of the Royal Astronomical Society},
  volume={518},
  number={3},
  pages={3966--3984},
  year={2023},
  publisher={Oxford University Press}
}

\clearpage

\begin{appendix}
\nolinenumbers

\onecolumn
\section{Uncertainties}
\label{app:substellar_radius}

Several important uncertainties in our simulations may have considerable effects on our results. As already discussed in \citet{mauchsorianoetal26-1}, the most important parameter determining the total occurrence of substellar companions to WDs is the age--metallicity relation, and due to variations of this relation in local volumes or the Milky Way, the fraction of WDs hosting planetary and brown dwarf companions may increase by a factor of $\sim2$. 

With regard to planetary survival, one of the most important physical properties determining the outcome is the binding energy of the stellar envelope. In addition to recombination energy, successive planetary engulfment events can contribute to envelope ejection, potentially allowing only the last planet to enter the envelope to survive \citep{Chamandy-21}. In this respect, our approach, which considers planetary systems with a single planet, may further underestimate the planetary survival fraction.

Another crucial parameter is the mass of the engulfed companion. The population synthesis of companions surviving CE presented here is based on the simulations by \citet{mauchsorianoetal26-1}, who assumed a companion mass distribution derived from observations of companions to FGK main-sequence stars \citep{cummingetal08-1,mayoretal11-1,adamsetal21-1}. This distribution originally extends up to $10$~\Mjup, but was extrapolated to larger companion masses, that is,  the brown dwarf regime. This probably leads to overestimating the occurrence of massive companions as we are extrapolating into the brown-dwarf desert. This uncertainty in our initial distribution should therefore translate to lower occurrence and detection rates.

Another uncertainty in our model is the treatment of stellar tides. 
We assumed the prescription by \citet{Villaver_2009} and \citet{Mustill_2012}. As discussed in detail in \citet{mauchsorianoetal26-1}, these tidal prescriptions are weaker than earlier suggestions by \citet{rasioetal96-1}. However, most likely even the prescription of \citet{Villaver_2009} overestimates the effects of stellar tides because of the assumed frequency dependence \citep{Esseldeurs2026}. The tidal prescription we used is therefore likely to generate more companions entering CE than a more realistic prescription. 

Finally, we assume a fixed radius of $R_p=1\,R_{\rm J}$ for determining survival, a reasonable approximation for brown dwarfs and giant planets in the mass range considered here \citep{Baraffe2003}. However, varying the radius between $0.8$--$1.2\,R_{\rm J}$, while keeping the stellar structure and orbital properties at CE onset fixed, changes the number of post-CE survivors, because $a_{\rm RLOF}$ scales directly with $R_p$. 
The survival fractions are therefore sensitive to this assumption, particularly for systems whose final separations lie close to the Roche-lobe overflow limit. At smaller adopted radii, a few massive companions also survive RGB or E-AGB engulfment, while survivors obtained with the $E_{\rm pot}+E_{\rm therm}$ prescription, as well as all survivors with $M_p<13\,M_{\rm J}$, remain associated with TP-AGB engulfment. Nevertheless, low-mass planets can still survive CE evolution throughout the explored radius range, while the predicted occurrence remains below $0.24\%$, preserving the expectation that surveys of WDs within 50 pc should yield few, if any, detections.

\end{appendix}

\end{document}